\documentclass[10pt,preprint]{aastex}

\begin{document}

\shortauthors{Luhman, McLeod, \& Goldenson}
\shorttitle{HST Companion Search in IC~348}

\title{An {\it HST} Search for Substellar Companions in the Young Cluster 
IC~348\altaffilmark{1}}

\altaffiltext{1}{
Based on observations made with the NASA/ESA Hubble Space
Telescope, obtained at the Space Telescope Science Institute, which is operated
by the Association of Universities for Research in Astronomy, Inc., under NASA
contract NAS 5-26555. These observations are associated with proposal ID 8573.}

\author{K. L. Luhman\altaffilmark{2,3}, K. K. McLeod\altaffilmark{4}, and 
N. Goldenson\altaffilmark{5}}

\altaffiltext{2}{Harvard-Smithsonian Center for Astrophysics, 60 Garden St.,
Cambridge, MA 02138; kluhman@cfa.harvard.edu.}

\altaffiltext{3}
{Visiting Astronomer at the Infrared Telescope Facility, which is operated
by the University of Hawaii under Cooperative Agreement no. NCC 5-538 with
the National Aeronautics and Space Administration, Office of Space Science,
Planetary Astronomy Program.}

\altaffiltext{4}{Whitin Observatory, Wellesley College, Wellesley, MA 02481;
kmcleod@wellesley.edu}

\altaffiltext{5}{Department of Astronomy, Wesleyan University, Wesleyan Station,
Middletown, CT 06459; ngoldenson@wesleyan.edu}

\begin{abstract}

We present the results of a search for substellar companions to members of
the star-forming cluster IC~348. Using the Wide Field Planetary Camera 2 
aboard the {\it Hubble Space Telescope}, we have obtained
deep, high-resolution images of the cluster through the F791W and F850LP 
filters. These data encompass 150 known members of IC~348, including 14 
primaries that are likely to be substellar ($M_1=0.015$-0.08~$M_\odot$).
The detection limits for companions to low-mass stars and brown dwarfs in the
PC images are $\Delta m_{791}=0$, 2.5, and 5.5 at separations 
of 0.05, 0.1, and $0\farcs3$, respectively, which correspond to 
$M_2/M_1=1$, 0.3, and 0.1 at 15, 30, and 90~AU.
Meanwhile, for heavily saturated solar-mass primaries in the WFC images, 
the limits are $\Delta m_{791}=0$ and 6 ($M_2/M_1=1$ and 0.04) at 0.2 and 
$0\farcs4$.  
The sky limiting magnitude of $m_{791}\sim26$ at large separations from
a primary corresponds to a mass of $\sim0.006$~$M_\odot$ according to the 
evolutionary models of Chabrier and Baraffe.
Point sources appearing near known and candidate cluster members are 
classified as either field stars or likely cluster members through their
positions on the color-magnitude diagram constructed from the WFPC2 photometry.
For the two faintest candidate companions appearing in these data, we have 
obtained 0.8-2.5~\micron\ spectra with SpeX at the IRTF. Through a comparison
to spectra of optically-classified dwarfs, giants, and pre-main-sequence 
objects, we classify these two sources as cluster members with spectral
types near M6, corresponding to masses of $\sim0.1$~$M_\odot$
with the models of Chabrier and Baraffe. Thus, no probable substellar 
companions are detected in this survey. 
After considering all potential binaries within our WFPC2 images,
we find that the frequencies of stellar and substellar companions within 
0.4-$5\arcsec$ (120-1600~AU) from low-mass stars ($M_1=0.08$-0.5) in IC~348
agree within the uncertainties with measurements in the field.
The factor of $\sim3$-10 deficiency in brown dwarfs relative to stars among 
companions at wide separations in IC~348 and across the much larger range of 
separations probed for field stars is equal within the uncertainties to the 
deficiency in brown dwarfs in measurements of mass functions of isolated 
objects. In other words, when defined relative to stars, the brown dwarf
``desert" among companions is also present among isolated objects, 
which is expected if stellar and substellar companions form in the same 
manner as their free-floating counterparts.
Meanwhile, among the 14 substellar primaries in our survey of IC~348,
no companions are detected.
This absence of wide binary brown dwarfs is statistically
consistent with the frequency of wide binary stars in IC~348.

\end{abstract}


\section{Introduction}

The multiplicity of stars and brown dwarfs is important in a variety 
of astrophysical contexts \citep{dm91}. 
The full characterization of multiplicity consists of the frequency of
secondary components as a function of several parameters, including primary
mass, secondary mass, mass ratio, separation,
eccentricity, age, and conditions of the birth place and subsequent environment.
A restricted portion of this phase space is probed by a typical survey
for binaries, which usually considers a discrete sample of primaries and 
employs a single method for identifying companions.
Radial velocity measurements have been used to search for close binaries
among stars in the local field \citep{al76,dm91,lat02} and in nearby open 
clusters \citep{gri88,mm99,aw99}.
In the past decade, the precision of these data has improved to the point of 
enabling the detection of companions below Jovian masses 
\citep{mq95,mb96,c97,noy97,but00,fis01,tin01}.
Meanwhile, direct imaging has been used to search for companions at 
wider separations near stars in the field 
\citep{nak95,reb98,sch00,bur00,opp01,giz01b,els01,kir01,wil01,hin02,pot02,lj02,mz04,mh04},
open clusters \citep{rg97,bou97,bou01,pat98,pat02}, 
young associations \citep{low99,low00,gue01,cha03}, and star-forming regions 
\citep{ghe93,ghe97,lei93,rz93,bra96,bra00,pro94,sim95,sim99a,sim99b,pet98,bk98,duc99,sca99}. 
Following the discovery of free-floating brown dwarfs \citep{shp94}, 
direct imaging has been extended to primaries with masses near and below the
hydrogen burning mass limit in the field
\citep{koe99,mar99,giz03,rei01,clo02a,clo02b,clo03,bou03,bur03,fre03,sie05},
open clusters \citep{mar98,mar00,mar03}, young associations \citep{cha04},
and star-forming regions \citep{neu02,bou04,luh04a}.

These multiplicity studies have produced a number of interesting results. 
The frequency of binaries appears to be significantly higher in Taurus 
\citep{ghe93,lei93,sim95,sim99a} and other low-density star-forming regions 
\citep{rz93,ghe97} than in denser star-forming clusters
\citep{pro94,pet98,duc99,sim99b}, open clusters \citep{rg97,bou97,bou01}, 
and the field \citep{dm91}.
The distribution of separations among multiple systems also varies with 
star-forming conditions \citep{bk98,sca99,pat02}.
In the distribution of companion masses for solar-type primaries, radial 
velocity surveys have revealed a dearth of brown dwarfs (20-80~$M_{\rm Jup}$) 
for separations less than 5~AU \citep{mb00}. This brown dwarf ``desert" 
extends to larger separations \citep{sch00,opp01,mz04}, except perhaps
beyond 1000~AU \citep{giz01a}. The extensive work on field dwarfs has revealed
progressively smaller binary fractions, smaller average
and maximum separations, and larger mass ratios with decreasing primary mass
from stars to brown dwarfs
\citep{dm91,fm92,rei01,pat02,bou03,bur03,clo03,giz03,sie05}.
No wide binary brown dwarfs ($a>20$~AU) have been found in these field 
surveys or in open clusters \citep{mar98,mar00,mar03}, 
while in star-forming regions one wide system has been recently discovered 
\citep{luh04a}.

We seek to improve constraints on the frequency of wide substellar companions 
to primaries from a solar mass down to below the hydrogen burning limit. 
Data of this kind can be 
obtained through deep high-resolution imaging of the members of nearby
young clusters, of which IC~348 is a prime example (2~Myr, 315~pc).
The low extinctions ($A_V=0$-4) toward most of the members of IC~348 allow for
observations at both optical and infrared (IR) wavelengths. In addition, because
the cluster is rich ($\sim400$~members) and compact ($D\sim30\arcmin$), 
multiple members can be observed simultaneously through wide-field 
high-resolution imaging. Motivated by these attractive characteristics
of IC~348, we have surveyed more than 100 of its known members for 
substellar companions using the Wide
Field Planetary Camera (WFPC2) aboard the {\it Hubble Space Telescope} ($HST$).
In this paper, we describe these observations (\S~\ref{sec:obs}) and 
the analysis of the resulting images (\S~\ref{sec:data}), identify candidate 
companions to known and suspected member of IC~348 (\S~\ref{sec:cand}), 
present near-IR spectra of the two faintest candidates (\S~\ref{sec:spex}), 
and discuss our data in the context of previous multiplicity measurements for
IC~348 (\S~\ref{sec:prev}) and for stars and brown dwarfs
in general (\S~\ref{sec:imp}).

\section{WFPC2 Observations}
\label{sec:obs}

\begin{figure}
\plotone{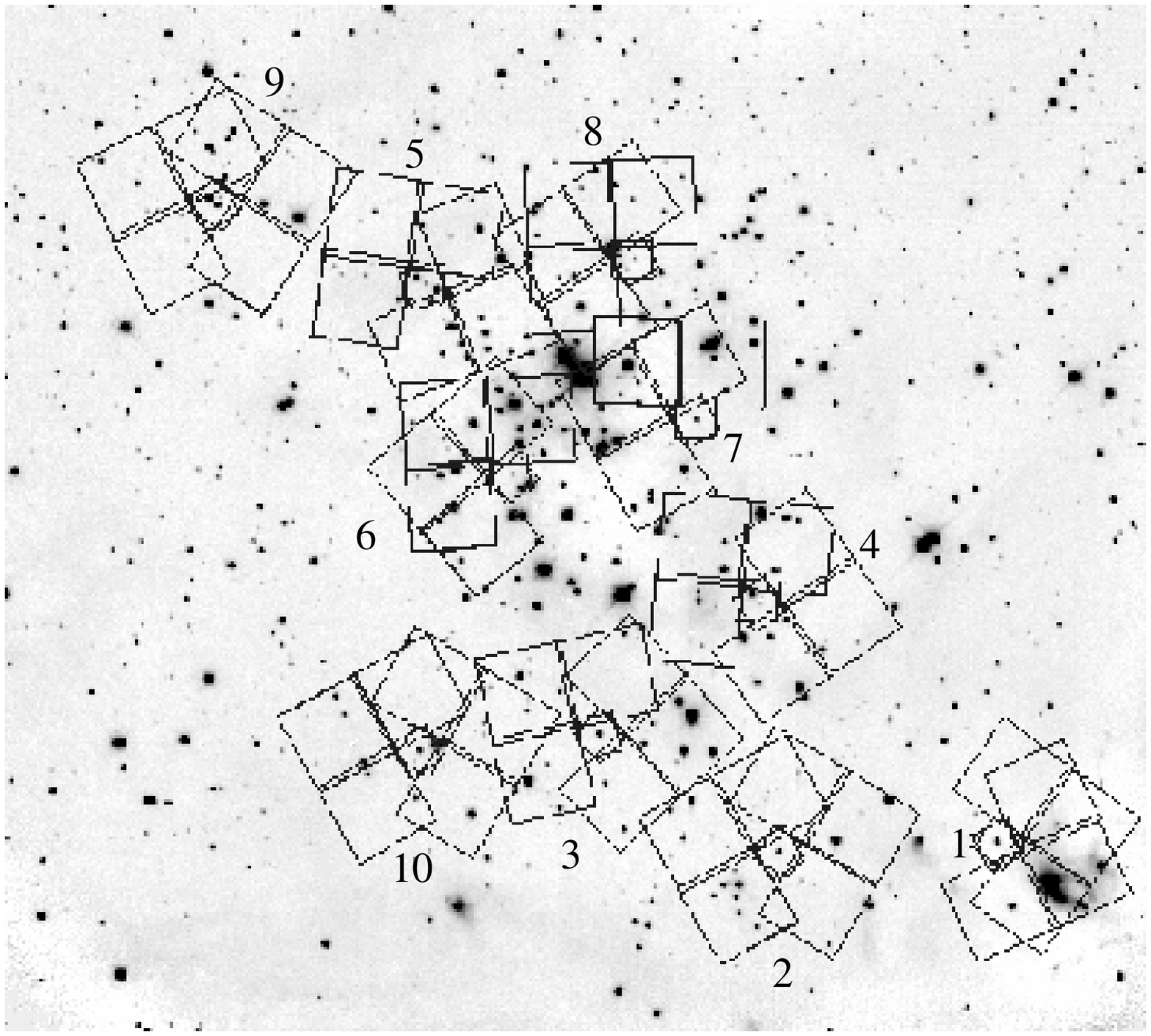}
\caption{
Fields toward the IC~348 cluster observed with WFPC2 on $HST$ shown with the 
$18\arcmin\times16\arcmin$ $H$-band image from \citet{mue03}. 
The field designations from Table~\ref{tab:log} are indicated.
}
\label{fig:fov}
\end{figure}

We used WFPC2 on $HST$ to obtain images of 10 positions in IC~348 during 40 
orbits from 2000 to 2002.  
The PC at each of the 10 positions was centered on a low-mass cluster
member (M5-M8, 0.15-0.03~$M_\odot$) from \citet{luh99}. These sources were
selected to be evenly distributed across the cluster to maximize the total 
number of cluster members appearing within all of the PC and WFC frames.
The initial 10 pointings are designated as POS1 through POS10.
We later observed each position again at a second spacecraft roll
angle to ensure the detection of any companions hidden by diffraction spikes
in the first orientation.
These rotated fields, designated POS1r through POS10r, also allowed
us to cover a greater area of the cluster. The 20 imaged fields are
shown in Figure~\ref{fig:fov} and are summarized in Table~\ref{tab:log}. 

Each of the 20 fields was observed in two consecutive orbits with the
filters F791W and F850LP. The system effective wavelengths with these
filters are 7900 and 9100~\AA, which are similar to the values of 
8100 and 9100~\AA\ for Cousins $I$ and SDSS $z\arcmin$ \citep{fuk96}.
The wavelengths of these WFPC2 bandpasses are long enough to provide good 
sensitivity 
to cool, red substellar objects. In addition, a color-magnitude diagram
constructed from photometry in these two bands is useful for separating 
cluster members (including resolved companions) from most field stars
(\S~\ref{sec:cand}).
In each filter, we obtained a pair of 400~sec exposures at each position
in a two-point diagonal dither pattern. Obtaining multiple exposures in
this way facilitated cosmic ray (CR) rejection.  We chose the dither 
spacing to be on half pixel centers to improve spatial sampling of the PSF. 
The targets centered on the PC, which have $I\gtrsim17$, were generally 
well-exposed without reaching saturation on the PC, although many of 
the brighter cluster members in the WFC frames were saturated. 
A gain of seven was used throughout the observations.

Detecting faint companions as close as possible to each primary 
requires accurate measurements of the complex PSF of $HST$. As described
in \S~\ref{sec:psfpc}, we have used a variety of observed and model stars to
accomplish this.  For one strategy, adopted because of its successful use
with quasar host images \citep{mm01}, we observed a PSF star at the center 
of the PC at the end of each orbit. These stars were selected within
$2\arcmin$ of the target, which was the maximum allowed slew that did
not require overhead for acquisition of new guide stars. The PSF stars
were observed in the same two-point dither pattern as our target
stars, although not with CR-split exposures. Typical exposures times
were 20 to 40 sec.

\section{WFPC2 Data Analysis}
\label{sec:data}

\subsection{Image Reduction}

\begin{figure}
\plotone{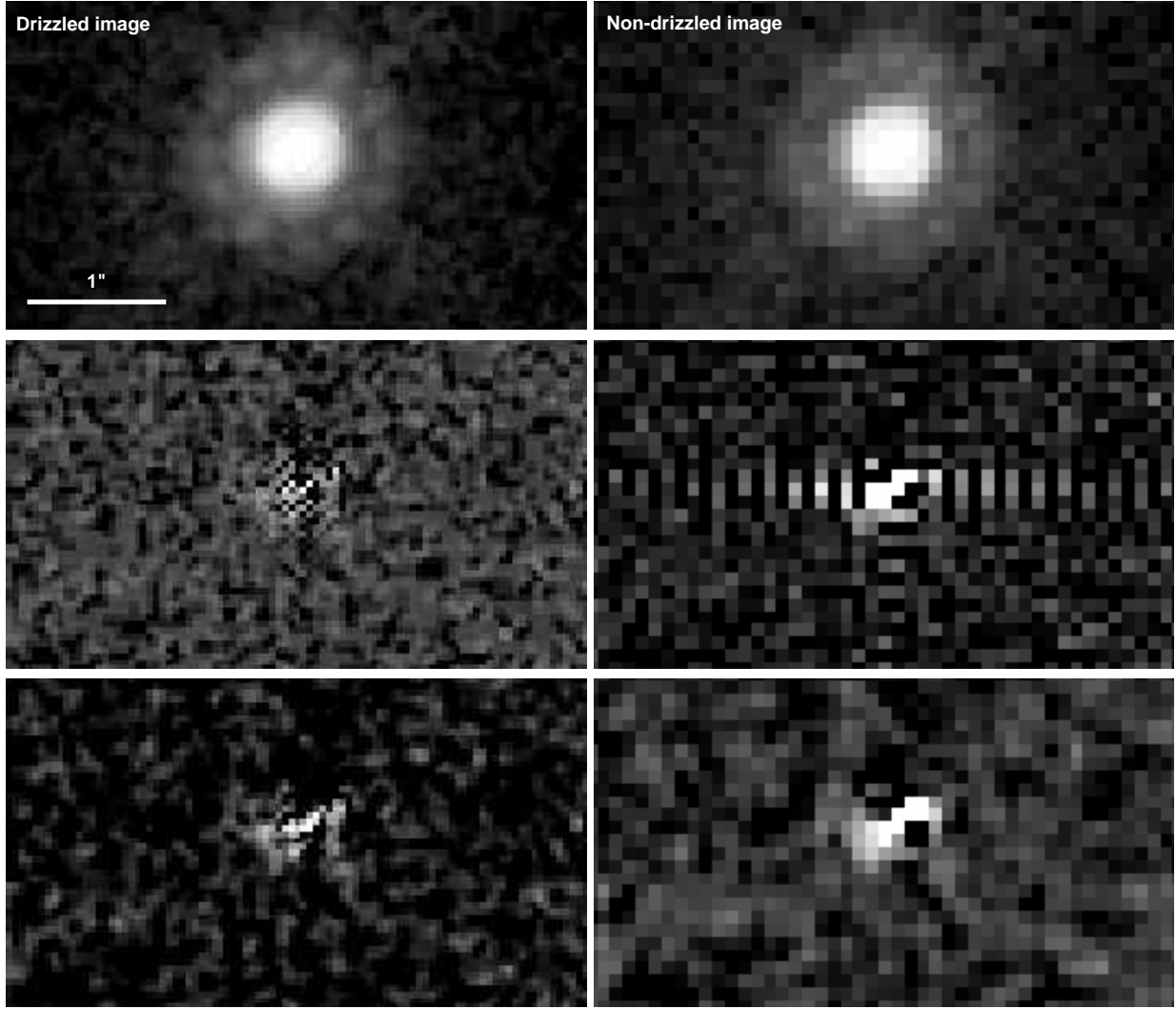}
\caption{
Improvement of PSF fitting due to drizzling.
{\it Top:} Drizzled and non-drizzled images of a star centered on the PC.  
{\it Middle:} Images after PSF subtraction. The ringing is an artifact of 
the Fourier transform used in the convolution.
{\it Bottom:} Images after PSF subtraction and smoothing. These data are shown 
on a logarithmic scale.}
\label{fig:drizzle}
\end{figure}

The WFPC2 images were reduced
with {\it IRAF\footnote{IRAF (Image Reduction and Analysis
Facility) is distributed by the National Optical Astronomy
Observatories, which are operated by the Association of Universities
for Research in Astronomy, Inc., under contract with the National
Science Foundation.}} and the astronomical imaging application {\it
ds9\footnote{ds9 is available at $http://hea-www.harvard.edu/RD/ds9$}}.
Starting with On-The-Fly-Calibrated images and bad pixel masks, we
performed cosmic ray rejection on image pairs to create a single image
for each dither position and filter. We measured the positional 
offsets of the dithers through cross-registration of the PC images.   
These offsets were combined with the known relative positions of the WFC and PC 
chips to arrive at the offsets among all of the dithered WFC and PC images.
To align the dithered images, we used the {\it STSDAS dither} package
and {\it drizzle} task \citep{fh02} to place the images and their associated 
bad pixel masks onto subsampled pixels. Optimal drizzle parameters were 
determined by experiments in which we applied our PSF-fitting technique as
illustrated in Figure~\ref{fig:drizzle} and measured the residuals of
the fit near the core of the star.  We found that a drop size of 0.6
pixels aligned on output pixel corners with a $2\times$ subsampled
pixel grid gave the best results. This combination is similar to, but
gives a somewhat better-resolved PSF than, a simple shift-and-add
procedure. During drizzling we applied the WFPC2 geometric distortion 
corrections provided by John Trauger and included in the package.
Finally, we combined the drizzled images to make a single image for each filter.

\subsection{Photometry and Astrometry}

We conducted a semiautomatic search for point sources in the PC and WFC frames. 
The objects were found by first combining the F791W and F850LP frames
into one lower-noise frame, and then running {\it daofind} with a 5~$\sigma$
threshold. Residual cosmic rays and point-like features in the PSFs of 
bright stars among these sources were rejected by visual inspection,
which also resulted in the identification of additional
objects that had been missed by {\it daofind}. We then performed
aperture photometry on unsaturated sources in the F791W and F850LP images 
and retained all objects with formal errors $<0.20~\rm mag$ in either filter.  
After including saturated stars, the resulting catalog contains 617 unique 
sources. According to the compilations of data for IC~348 from \citet{luh03} 
and \citet{luh05}, our WFPC2 catalog contains 150 known cluster members, 
which are listed in Table~\ref{tab:mem}.
Coordinates for all sources were determined from the F791W images, and then 
transformed to the astrometric system defined by stars in the optical 
images from \citet{luh03}, whose plate solutions were originally derived 
from 2MASS astrometry.
The coordinates of unsaturated and saturated sources have precisions of 
$\sim0.002$ and $0\farcs02$, respectively. 

The aperture photometry was performed with a radius of four pixels in the final 
subsampled images, which translates to $r\approx0\farcs1$
and $0\farcs2$ on the PC and WFC, respectively.  We determined
aperture corrections to $r=0\farcs5$ for each chip+filter combination
through measurements of well-exposed 
stars and found no significant trend with time or position on the arrays.
The aperture corrections, which ranged from $0.19-0.27~\rm mag$ on the WFC
chips and $0.58-0.62~\rm mag$ on the PC, added $0\farcs05$ mag to the
photometric uncertainty.  Experiments with model PSFs using {\it TinyTim
v.\ 6.0}~\citep{kh01} indicated that any additional aperture corrections
for star color are smaller.

Charge transfer efficiency (CTE)
corrections were computed following \cite{dol00}\footnote{Used May 2002
simple formula from $http://www.noao.edu/staff/dolphin/wfpc2_calib/$}.
The correction accounts for position on the chip, epoch of the
observations, background level, and source flux.
Typical CTE corrections were in the range $0.05-0.25$ and $0.05-0.45~\rm mag$
in F791W and F850LP, respectively, but for a few faint sources
near the chip edges in F850LP they exceeded a factor of two.  We have
not added uncertainty from the CTE correction. Comparing 
the formal errors we quote with the photometric scatter for the nearly
200 sources that were observed in two separate pointings, and hence on
different parts of the chips, we found that any additional
uncertainties must be negligible. 

Magnitudes were converted to STMAG infinite aperture 
magnitudes through an additional $0.1$ mag correction and application
of the photometric zero points from the image headers.
The photometry for the unsaturated known members of IC~348 within
our WFPC2 frames are provided in Table~\ref{tab:mem}.

\subsection{PSF Analysis for the PC Images}
\label{sec:psfpc}

We used a variety of techniques to perform PSF removal for the
known cluster members centered on the PC frames,
and for other stars found in those images. The PSF-subtraced images of 
these objects were then visually inspected for companions.
For the PSF removal, we considered a suite of PSFs,
including the PSF star observed at the end of the orbit,
PC targets from other orbits, which were taken at different times but
which generally have 
higher signal-to-noise than the PSF stars, and model WFPC2 PSFs
generated by TinyTim.  The latter were generated with $2\times$
subsampling and were convolved with the scattering function given in
the TinyTim manual.  We adopted a TinyTim spectral type of M1.5 to
represent the relatively red colors of our stars.  Experiments with
other spectral types showed negligible differences within the scope of
our fitting procedure.

\begin{figure}
\plotone{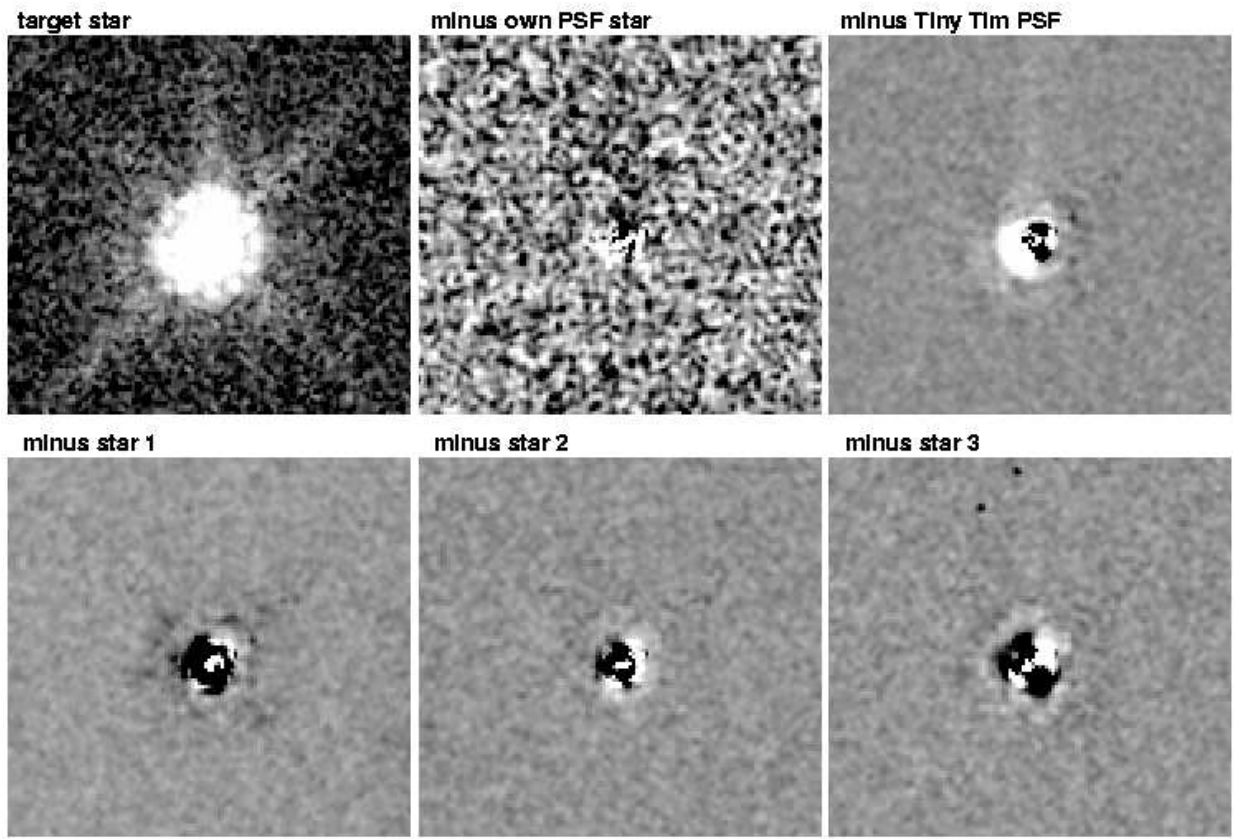}
\caption{
Examples of PSF fits to PC targets. 
{\it Top left:} The unsubtracted PC image of an unsaturated star.
{\it Top middle:} The PSF star observed on the same orbit provides a good 
match in the core, but scaled-up background noise is high in the wings.
{\it Top right:} The fit with a TinyTim PSF is relatively poor.
{\it Bottom:} The PSFs of stars in PC images on other orbits 
(stars 1-3) provide fits that are good in the wings but varying in quality 
in the core.
}
\label{fig:imfits}
\end{figure}

We determined best-fit models for each PC target by convolving a
point source with the PSF, and then varying the parameters to 
minimize the sum of the squares of the residuals over all 
the pixels \citep{leh00}. We excluded saturated pixels and cosmic rays
through the use of a mask. The position and magnitude of the point source 
and the level of the background were allowed to vary.
When using TinyTim PSFs or star images not taken on the same orbit as the 
target, we also allowed the PSF to ``blur'' by smoothing with a
2-dimensional circular Gaussian whose width is an additional
parameter of the fit. Figure~\ref{fig:imfits} provides
an example of the range of residuals from the various fits.  
These results were typical of unsaturated stars in the PC images 
but better than those achieved for stars that were strongly saturated
or near the edges of the PC frames.

As expected, the PSF star for a given target provided a superior fit in the 
center, allowing us to search for companions very near the core. However,
the residuals outside the core were large owing to the amplification of
noise in the wings that results from scaling up the image of the PSF star,
which usually exhibited lower signal-to-noise than the target. Thus, to search 
for companions in the wings, we relied instead on the fits to PSFs from
other target stars. Although the TinyTim PSFs are noiseless, they
generally provided worse fits, especially in the core.

In addition to PSF fits, we performed a Lucy-Richardson
deconvolution on each PC star with the {\it IRAF lucy} routine.
Observed PSFs were too noisy for restoration; TinyTim PSFs worked
somewhat better. We also examined the radial profiles to
search for close companions of nearly equal magnitudes, which would manifest 
themselves as a slight elongation of the stellar core. This analysis was
sensitive to equal magnitude binaries with separations down to $0\farcs05$.

To test our sensitivity to companions of various magnitudes, we
generated a suite of artificial companions (point sources convolved with the
star image) at different separations and azimuths from a typical
target star.  We added them to the star image and then performed our 
fits, deconvolutions, and radial profile analysis on the resulting images.
We carried out experiments on two stars that encompass the (small)
range of brightness of our PC targets. The results are shown in
Figure~\ref{fig:sensitivity}. 

In the core ($r<0\farcs15$), the best sensitivity to companions was produced by 
the fits with the target's own PSF star and from the radial profile analysis.
At intermediate separations ($0\farcs15<r<0\farcs5$), the PSF has a ring of 
point-like sources that must be subtracted cleanly to reveal any faint 
companions. In this regime fits using other of our PC target stars as 
PSFs produced the best results. Because of time-dependent and unpredictable 
PSF variations, we executed the fits with an ensemble of other target stars
to find a suitable one.  
At wide separations ($r>0\farcs5$), simple visual inspection of the 
unsubtracted images was sufficient to identify companion point sources.

\subsection{PSF Analysis for the WFC Images}
\label{sec:psfwfc}

In addition to the cluster members in the PC images, more than 100 known 
members appeared in the data from the flanking WFC chips.
The WFC data also included two dozen sources that are candidate members 
based on the color-magnitude diagrams from \citet{luh03} and this work.
We have applied our PSF analysis to these WFC sources, with some differences
from the procedure described in the previous section.
First, PSF star observations at the chip positions of the WFC sources, which
were scattered across the chips, were not available. Therefore, we
used only the model PSFs generated by TinyTim for the appropriate
chip and position of each star. Second, most of the cluster members in the WFC
frames were saturated. Although we masked the saturated cores during the 
fitting procedure, the fits were often very poor because of the limitations 
of the models and the extreme instability of the high spatial frequency 
features in the PSF wings.
To first order, the detection limit for companion point sources as a 
function of separation from a unsaturated or lightly saturated WFC star 
resembles those from the PC data, except at double the spatial scale 
within $0\farcs2$ owing to the size of the WFC pixels.  
However, for the heavily saturated stars, the sensitivity is worse and 
is not easily quantified because of the asymmetrical nature of the areas
subject to bleeding. As an example, in Figure~\ref{fig:sensitivity} we
show the detection limit for position angles free of bleeding around 
one of the brighter cluster members in the WFC frames, source 37 ($I=13.2$). 
For the angles containing severe bleeding, the limit is worse, with 
$\Delta m=2$ moving out to $\sim0\farcs4$.

\section{Identification of Candidate Companions}
\label{sec:cand}

To identify potential companions within our catalog of WFPC2 sources, we first 
describe a means of discriminating between probable field stars and candidate 
members of the IC~348 cluster through a color-magnitude diagram constructed
from the WFPC2 photometry. We then select all WFPC2 sources that are 
within $5\arcsec$ of known and candidate members and evaluate the status
of these candidate companions.

A point source detected at a small projected separation from a member of
the IC~348 cluster could be a companion, an unrelated cluster member, or a
field star in the foreground or the background of the cluster. 
To distinguish the third possibility from the first two,
we can make use of the fact that most field stars exhibit optical colors 
and magnitudes that are distinct from those of members of a young nearby 
cluster like IC~348. In Figure~\ref{fig:iz}, we plot a color-magnitude 
diagram consisting of photometry at F791W and F850LP for all unsaturated 
sources in the WFPC2 images, except for known foreground and background stars 
compiled by \citet{luh03}. We define a boundary below the lower envelope
of the sequence of known members of IC~348 to separate candidate cluster
members and probable field stars. For a variety of photometric systems, 
the $I-Z$ colors of late-type field dwarfs are roughly constant from M8 
to mid-L \citep{sh00,dah02,dob02}. Because the $m_{791}-m_{850}$ colors for 
WFPC2 could behave in the same manner, below the end of the 
sequence of known members we define the boundary to be vertical.
Previously unclassified sources appearing above this boundary are candidate
members of the cluster, while the objects below it are likely to be field stars.
Note that the four known members below the boundary in 
Figure~\ref{fig:iz} (435, 622, 725, 1434) are also subluminous in 
previously published optical color-magnitude diagrams, which may 
indicate that they are seen primarily in scattered light \citep{luh03}.
Indeed, one of these objects, 1434, appears slightly extended in the WFC 
images.

For each known member in the WPFC2 images (Table~\ref{tab:mem}) 
and each object that is a 
candidate member according to Figure~\ref{fig:iz}, we have searched the 
WFPC2 catalog for sources that are within a projected separation of $5\arcsec$,
which corresponds to $\sim1600$~AU at the distance of IC~348.  
We selected $5\arcsec$ as the outer limit because the 
expected number of chance alignments of unrelated cluster members 
becomes significant ($\gtrsim10$) beyond this distance.
We exclude pairs in which one of the stars has been classified as a field 
star in previous studies.
For the resulting 31 pairs, astrometry and photometry are listed in 
Table~\ref{tab:pairs} and F791W images are shown in Figure~\ref{fig:mempairs5}.
We also provide in Table~\ref{tab:pairs} the status of each object as a 
member of IC~348, candidate member, or probable field star, which was 
assigned in the following manner. Individual stars that have been classified 
as members of IC~348 through previous work \citep[see][]{luh03} are listed 
as such. In addition, if a tight equal-magnitude pair ($<1\arcsec$) exhibits 
evidence of membership in unresolved observations of the two stars 
(e.g., 1A and 1B), we take each star to be a cluster member. 
The remaining stars are classified as candidate members or field stars 
based on their positions in Figure~\ref{fig:iz}, where we indicate 
all of the unsaturated stars within the 31 pairs. 
These pairs consist of 36 known members,
16 probable field stars, six candidate members, and one star that 
cannot be classified because it is detected only at F791W.
If each pair containing a field star is omitted, we arrive at 14 remaining
pairs that are potential binary systems, nine of which consist of two known 
members and five of which consist of a known member and a candidate member.
For each of these five candidate members, 
we estimated a mass by assuming that it has the same reddening and age
as the known member with which it is paired and that the ratio of their 
luminosities is equal to the ratios of the fluxes at F791W. 
We then combined these assumptions with the evolutionary models of \citet{bar98}
and \citet{cha00} to arrive at the mass estimates in Table~\ref{tab:pairs}. 
If the five candidates are bona fide cluster members, this analysis 
indicates that they are probably low-mass stars at 0.1-0.2~$M_\odot$. 
Masses just below the hydrogen burning mass limit are possible but appear 
less likely.
Among the pairs of known members, source 761 is a brown dwarf according to 
\citet{luh03}. 
To establish the cluster membership of the candidates, additional observations 
are required, such as spectroscopy in which the components of the pairs
are resolved. In the next section, we present data of this kind for the 
two faintest candidates.

Are the 14 pairs of known and candidate members in Table~\ref{tab:pairs} true 
binary systems or unrelated stars seen in projection near each other?
We addressed this question by performing a Monte Carlo simulation of the 
projected separations of unrelated cluster members in IC~348. 
We began by measuring the surface density of the 267 known members within
the $16\arcmin\times14\arcmin$ area considered by \citet{luh03}. The current
census of cluster members is nearly complete for this field and most of 
our WFPC2 pointings fall within it.
In a given realization of the simulation, 2-dimensional positions of 267 stars 
were randomly drawn from the measured surface density distribution. The 
projected separation to the nearest star was computed for each cluster member. 
This model indicates a 90\% probability of 0-2 and 3-11 unrelated pairs 
with projected separations of $\leq2\arcsec$ and 2-$5\arcsec$, respectively.
Because our WFPC2 images contain $\sim150$ known members of IC~348, 
these expected numbers of unrelated pairs should be reduced by a factor of
1.8 for comparison to the pairs observed by WFPC2. 
Therefore, among the
14 potential binaries in Table~\ref{tab:pairs}, 8-9 of the 9 pairs 
with separations $\leq2\arcsec$ are probably true binaries, while 2-5 of the 
5 pairs at 2-$5\arcsec$ probably contain unrelated cluster members.
To test definitively whether these pairs are true binaries, common proper
motions could in principle be used. However, such measurements
would require extremely high precision, because even unrelated
members of the cluster share the same motion to within a 
few~km~s$^{-1}$, or $\sim0\farcs001$~yr$^{-1}$.

Finally, we point out a few interesting sources from the WFPC2 images 
in addition to the candidate companions. 
One of the six candidate members within the pairs in Table~\ref{tab:pairs}, 
source 596, is not a potential companion because it is paired with a likely 
field star. However, based on its position in Figure~\ref{fig:iz}, this
object could be a free-floating brown dwarf with an extremely low mass
(5-10~$M_{\rm Jup}$) if it is a member of IC~348. 
The low-mass members 906 (M8.25) and 1434 (M6) appear to be slightly 
extended in the WFC images, which could reflect the presence of close 
binaries or resolved circumstellar material.

\section{Spectroscopy of Candidate Companions}
\label{sec:spex}

We obtained near-IR spectra of the two faintest candidate companions 
identified in the previous section, sources 78B and 166B.
These data were collected with the spectrometer SpeX \citep{ray03} at the 
NASA Infrared Telescope Facility (IRTF) on the night of 2004 November 12.
The instrument was operated in the prism mode with a $0\farcs8$ slit,
producing a wavelength coverage of 0.8-2.5~\micron\ and a resolution of
$R\sim100$. For each candidate companion, the slit was placed along the axis 
connecting the candidate and its primary. 
In the resulting images, the components exhibited FWHM$=0\farcs75$, and thus
were sufficiently resolved from each other for the extraction of separate
spectra. The spectra were reduced with the Spextool package \citep{cus04}. 
We selected aperture radii of $0\farcs5$ and $0\farcs4$ for 78B and 166B,
respectively.
To correct for contamination by the primaries, the same aperture on the 
opposite side of each primary was used for background subtraction.
The data were corrected for telluric absorption with the method described 
by \citet{vac03}. For comparison, we also obtained spectra of several 
optically-classified late-type dwarfs, giants, and pre-main-sequence objects.

To determine if 78B and 166B are members of the young cluster IC~348
rather than field dwarfs or giants, we can examine spectral features
that are sensitive to surface gravity. 
In Figure~\ref{fig:spex1}, 
we compare the spectrum of 78B to data for field M dwarfs and giants.
We show only 78B because the spectra of 78B and 166B are very similar.
The dwarfs are Gl~406 and LHS~2065 \citep[M6V and M9V,][]{kir91,hks94} and 
the giants are VY~Peg and IRAS~09540-0946 \citep[M7III and M8III,][]{kir97}
\footnote{In contrast to the published optical spectral types, IRAS~09540-0946 
is earlier than VY~Peg and appears to be mid-M in our IR spectra. 
Variability in spectral type of this kind is often observed in M giants.}
The broad plateaus in the $H$ and $K$ spectra of the field dwarfs and giants
are absent in 78B and 166B, which instead exhibit sharply peaked, 
triangular continua, as previously observed for young late-type objects
\citep{luc01}. These two objects also lack strong Na absorption at 
2.2~\micron\ and strong CO absorption at 2.3~\micron, further indicating
that they are not dwarfs or giants, respectively. In general, the spectra
of 78B and 166B are not well-matched by those of the field dwarfs and giants.
Meanwhile, the gravity-sensitive spectral features of 78B and 166B agree well 
with those of late-type pre-main-sequence objects in Taurus, as shown in 
Figure~\ref{fig:spex2}. The significant reddening toward 78B and 166B is
independent evidence that they are not foreground objects. They are also too 
bright to be background field dwarfs, which appear well below the sequence 
of cluster members on the color-magnitude diagram in Figure~\ref{fig:iz}. 
Therefore, we conclude that 78B and 166B are not field dwarfs or giants 
and are instead young members of the IC~348 cluster. 

The spectra of 78B and 166B are similar in terms of both the overall slope 
and steam band 
strengths, which imply similar reddenings and spectral types, respectively.
After comparing these spectra to data for optically-classified young 
sources at various spectral types, we find that 78B and 166B are best fit by 
V410~X-ray~3 \citep[M6,][]{ss94,luh98a,bri98} and 
V410~Anon~13 \citep[M5.75,][]{bri02}, respectively, as shown in 
Figure~\ref{fig:spex2}. If the intrinsic 0.8-2.5~\micron\ 
slope of young objects are similar to those of dwarfs, then a comparison 
to the M6 field dwarf Gl~406 indicates reddenings of $A_V=5.5\pm0.5$ for
both 78B and 166B. When combined with the temperature scale of \citet{luh03}
and the evolutionary models of \citet{bar98} and \citet{cha00}, 
a spectral type of M6 corresponds to a mass of $\sim0.1$~$M_\odot$. 

\section{Comparison to Previous Work in IC~348}
\label{sec:prev}

Several of the pairs listed in Table~\ref{tab:pairs} have been resolved
in previous imaging surveys of IC~348. For instance, the candidate companion
78B was first detected by \citet{her98} and subsequently classified as
M6.8 -- and thus a possible brown dwarf -- through narrowband photometry of
near-IR steam absorption by \citet{ntc}.
However, through the near-IR spectroscopy in the previous section,
we measured a slightly earlier spectral type of M6 for this object, 
which implies a higher mass, probably just above the hydrogen burning limit
according to the temperature scale of \citet{luh03} and the evolutionary 
models of \citet{bar98} and \citet{cha00}. This difference in spectral types is
consistent with the systematic errors in the steam classifications 
of \citet{ntc} that were noted by \citet{luh03}.
Whereas 78B and some of the other wide pairs in IC~348 were found 
serendipitously during work on the cluster's stellar population,
a dedicated multiplicity survey of IC~348 was performed by 
\citet{duc99} using near-IR adaptive optics imaging.
Their targets consisted of 75 systems identified as cluster members by 
\citet{her98} and \citet{luh98b} and 12 additional primaries from \citet{her98}
that lacked membership information. 
In the remainder of this section, we update the analysis of 
\citet{duc99} to incorporate the latest membership data \citep{luh03,luh05} 
and the WFPC2 images from this work.

In the tabulation of the results of their multiplicity survey, \citet{duc99} 
presented 14 probable binaries, five pairs containing a known or likely  
field star, and 68 unresolved stars. 
Seven of the 14 probable binaries are in Table~\ref{tab:pairs}, which include
1AB, 9AB, 12AB, 52/30095, 92AB, 99AB, and 226/30114. 
We classified all of the components of these seven pairs as known or candidate
cluster members. Six of these pairs have separations less than $2\arcsec$ and
therefore are probably true binaries on the basis of the Monte Carlo 
simulation from the previous section, while the binarity of 92A/B is less
certain given its relative wide separation of $3\farcs13$.
Another three of the 14 pairs, 10/45, 16/33, and 221 
\citep[identifications from][]{luh03},
are within our WFPC2 images but not listed in Table~\ref{tab:pairs}.
The first two pairs have separations of $6\farcs46$ and $5\farcs41$, 
respectively, which are beyond the $5\arcsec$ limit that we consider.
Source 221 is unresolved in our WFC images, which is consistent with
the $0\farcs13$ separation measured by \citet{duc99}.
Among the five pairs with known or probable field stars in \citet{duc99},
IfA~100, 124, and 137 are within the WFPC2 fields. IfA~100 appears to 
be a field star based on the absence of near-IR steam absorption in the
data from \citet{ntc}
and its position in color-magnitude diagrams from \citet{luh03}, in agreement
with the classification of \citet{duc99}. 
They also suggested that the faint sources detected near 
IfA~124 and 137 were background stars rather than cluster members, which
is supported by the data from \citet{ntc} and the positions
of these sources in our color-magnitude diagram in Figure~\ref{fig:iz}.
Finally, the 68 unresolved stars from \citet{duc99} consist of 58 known
members, six known field stars (IfA 67, 89, 147, 165, 193, 210),
four probable field stars according to the color-magnitude diagrams from 
\citet{luh03} (IfA 41, 61, 67, 255). Our WFPC2 images contain 35 of the
68 unresolved stars. We detect faint objects near three of them (15, 266, 165), 
all of which are probable field stars based on Figure~\ref{fig:iz}.

\citet{duc99} computed a binary fraction for the 66 systems in their sample
that were taken from \citet{her98}, 12 of which exhibited likely companions
in their data. As shown above, our multiplicity results agree with those
from \citet{duc99} for that sample. The only necessary revision to their 
estimate is the removal of two foreground stars, IfA 89 and 210 
\citep{fre56,sch99}, from the unresolved sources in their sample, which 
has negligible effect on their calculation.
\citet{duc99} found that their estimate of the binary fraction for this 
sample in IC~348 was consistent with that of field G and M dwarfs for the
same range of separations, 0.1-$8\arcsec$. 
They also noted an absence of brown dwarf companions in their data.
The results of our survey are consistent with the low frequency of wide
substellar companions from \citet{duc99}. 

\section{Implications of Companion Search}
\label{sec:imp}

In this section, we first describe the detection limits for companions 
in our WFPC2 images in terms of mass. We then combine these limits and the 
candidate companions appearing in our data to constrain the multiplicity of the 
solar-mass stars, low-mass stars, and brown dwarfs in IC~348. 
We finally comment on the brown dwarf desert in the context of these data
for IC~348 and measurements of initial mass functions of isolated stars
and brown dwarfs.

\subsection{Mass Detection Limits}

\begin{figure}
\plotone{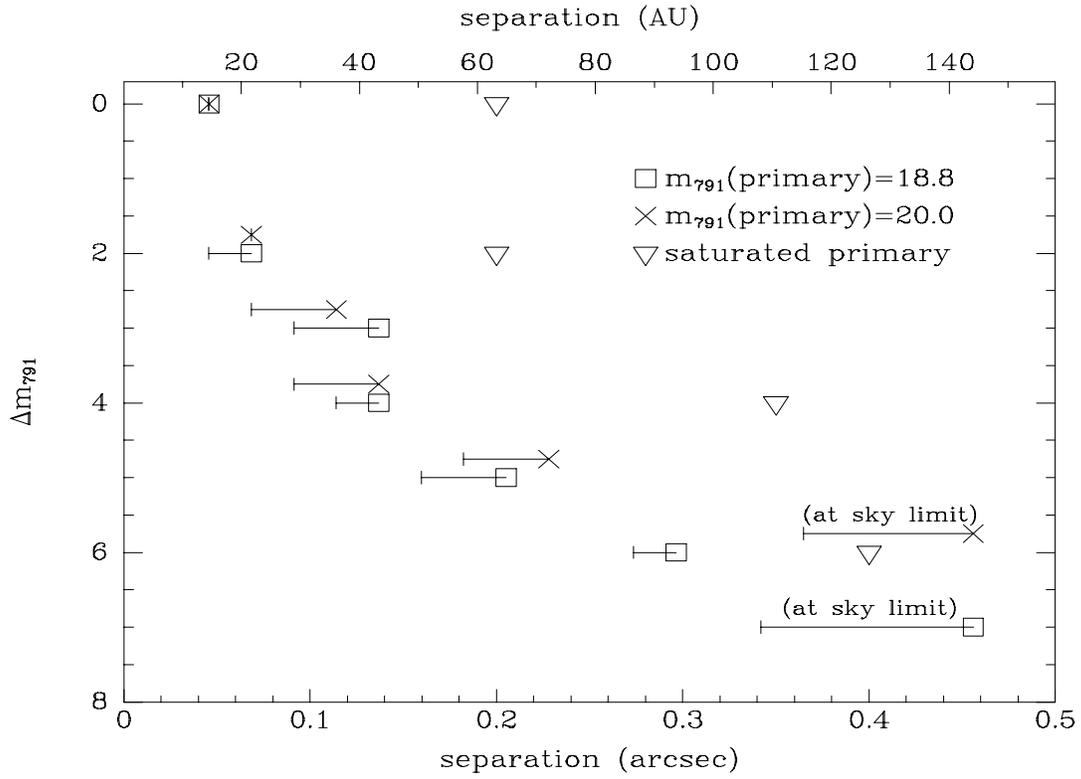}
\caption{
Detection limits for point sources near two unsaturated stars in
the PC images ({\it boxes and crosses}) and a heavily saturated star in
the WFC images ({\it triangles}) as a function of angular separation and after 
subtracting the F791W magnitude of the primaries. 
}
\label{fig:sensitivity}
\end{figure}

\begin{figure}
\plotone{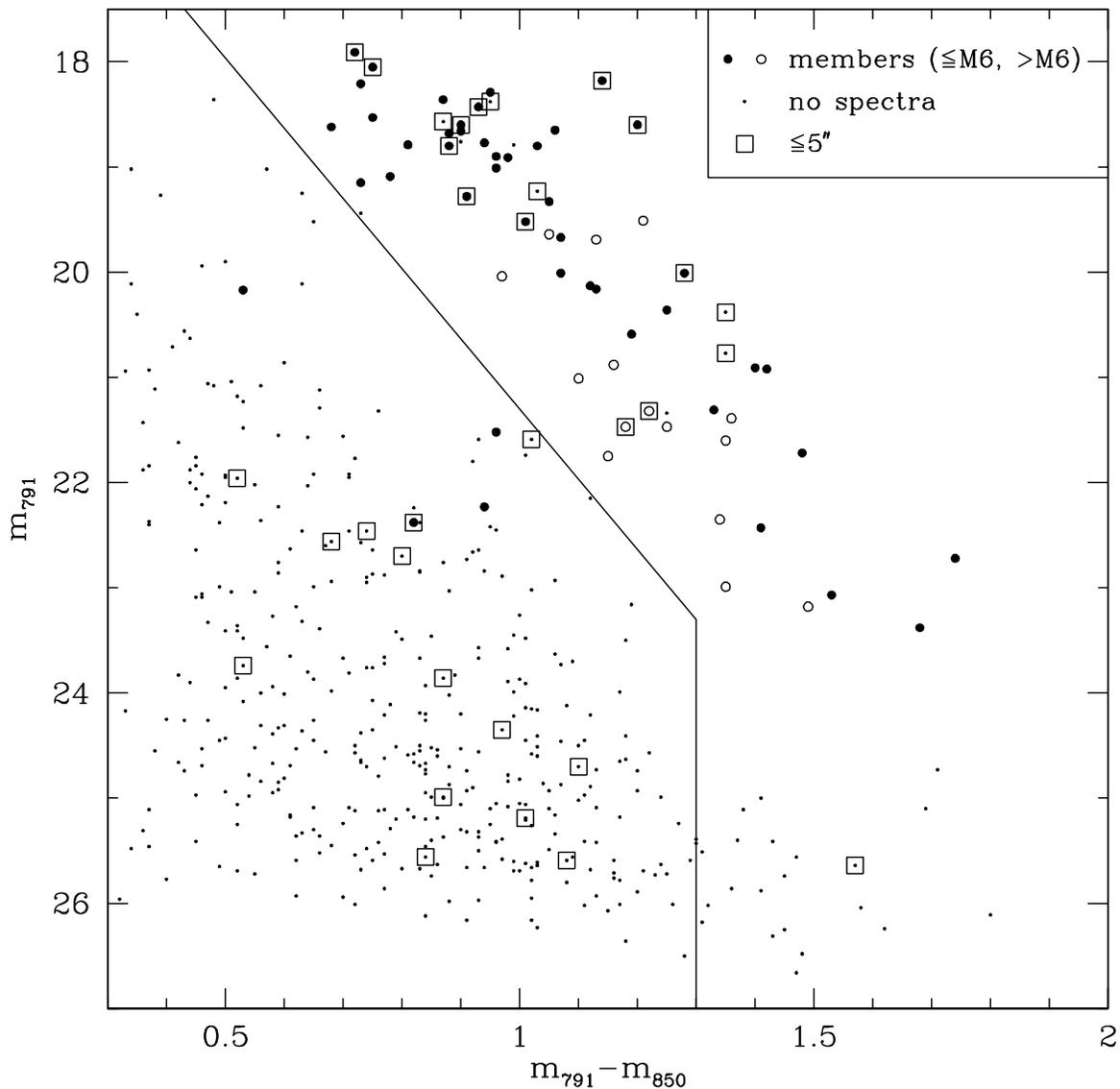}
\caption{
Color-magnitude diagram for unsaturated stars in the WFPC2 frames of the 
IC~348 cluster. 
We show the known cluster members at $\leq$M6 and $>$M6 ({\it large points and
circles}) while omitting known field stars. 
Members with spectral types later than M6 are likely to be brown dwarfs. 
The solid boundary was designed to follow the lower envelope of the sequence
of known members. 
Among the remaining sources that lack spectroscopic
measurements ({\it small points}), stars that are above and below this boundary
are candidate members and likely field stars, respectively.
We indicate the components of pairs with separations less than $5\arcsec$ 
in which at least one star is a known member or a candidate member 
({\it squares}). Near-IR spectra for the two candidate companions near 
$m_{791}=20.5$ and $m_{791}-m_{850}=1.35$, 78B and 166B, are shown in 
Figs.~\ref{fig:spex1} and \ref{fig:spex2}.}
\label{fig:iz}
\end{figure}

\begin{figure}
\plotone{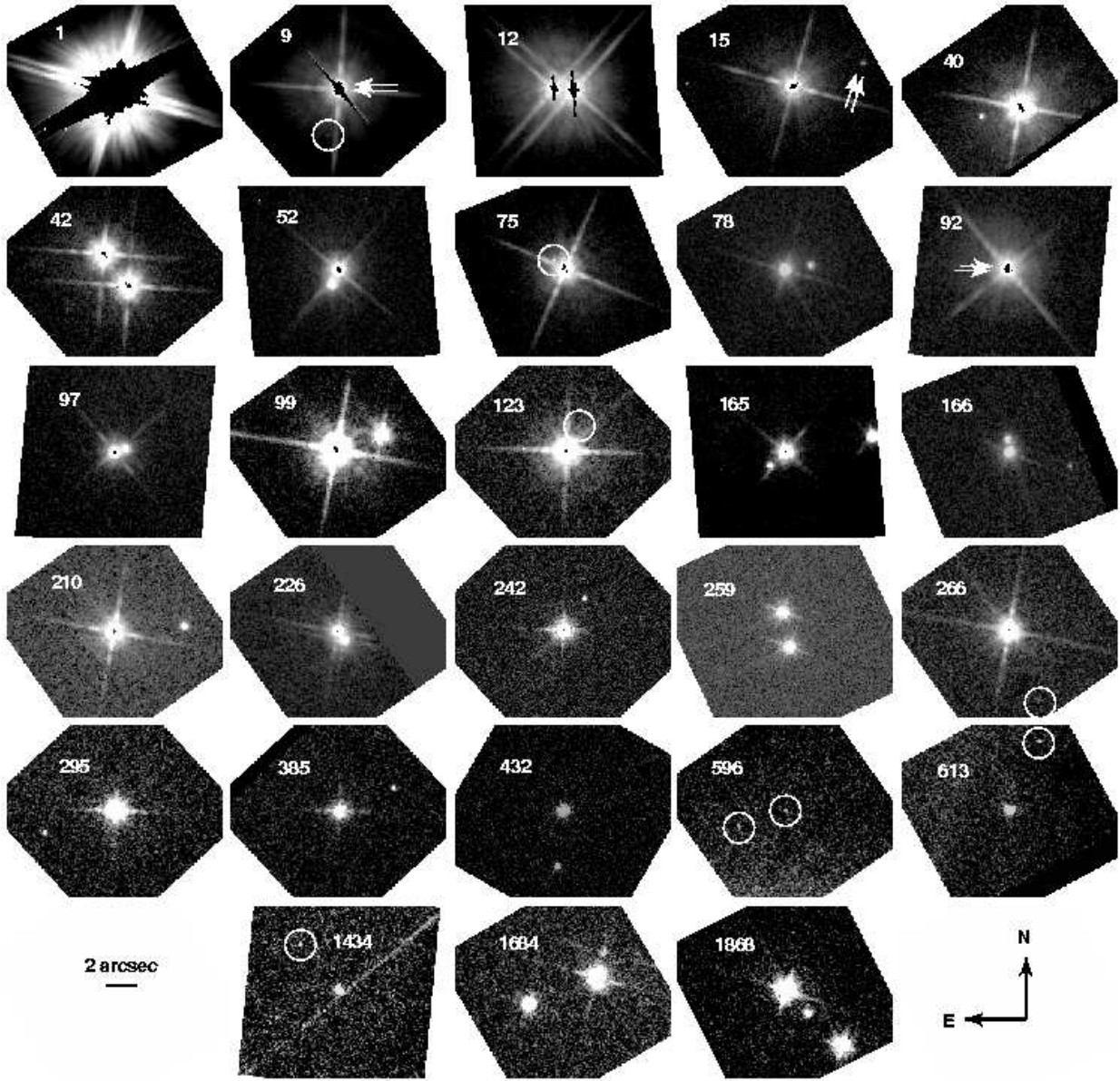}
\caption{
WFPC2 F791W images of pairs with separations less than $5\arcsec$ 
containing at least one known or candidate member of IC~348.
The status of each object as a known member, candidate member, or 
probable field star is indicated in Table~\ref{tab:pairs}.
Near-IR spectra of the candidate companions to 78 and 166 are shown
in Figs.~\ref{fig:spex1} and \ref{fig:spex2}.
These images have dimensions of approximately $11\arcsec\times11\arcsec$.
}
\label{fig:mempairs5}
\end{figure}

\begin{figure}
\plotone{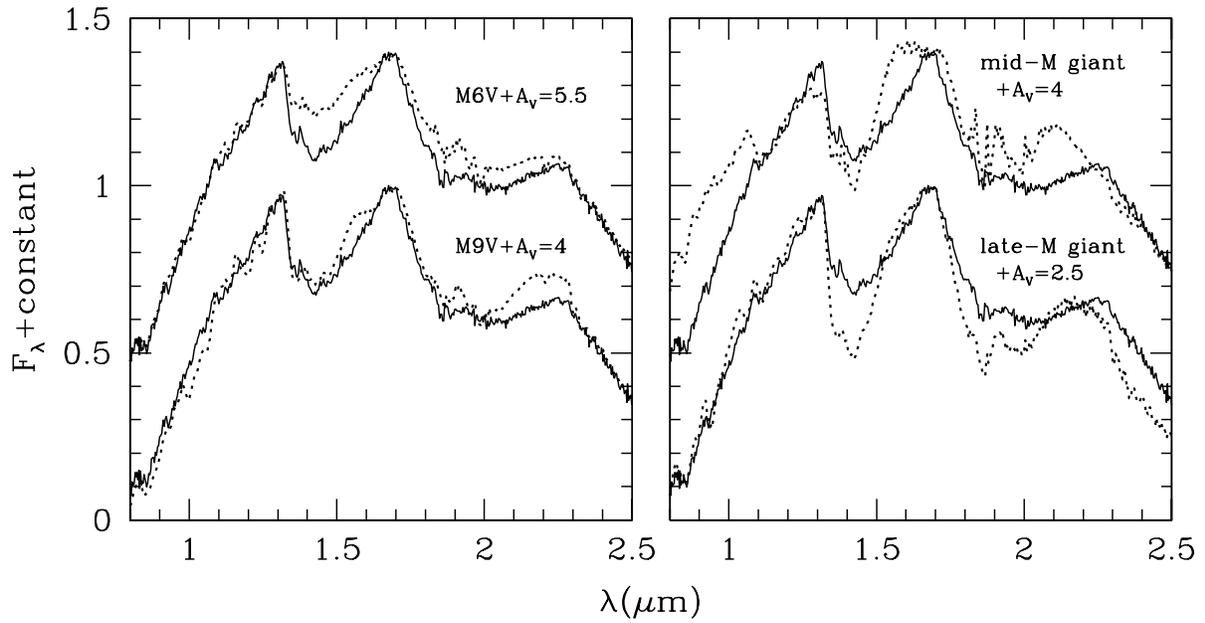}
\caption{
Near-IR spectrum of candidate companion 78B compared to spectra of field
dwarfs ({\it left}) and giants ({\it right}).
The latter have been reddened to match the overall slope of the spectrum of 78B.
The spectra have a resolution of $R=100$ and are normalized at 1.68~\micron.
}
\label{fig:spex1}
\end{figure}

\begin{figure}
\plotone{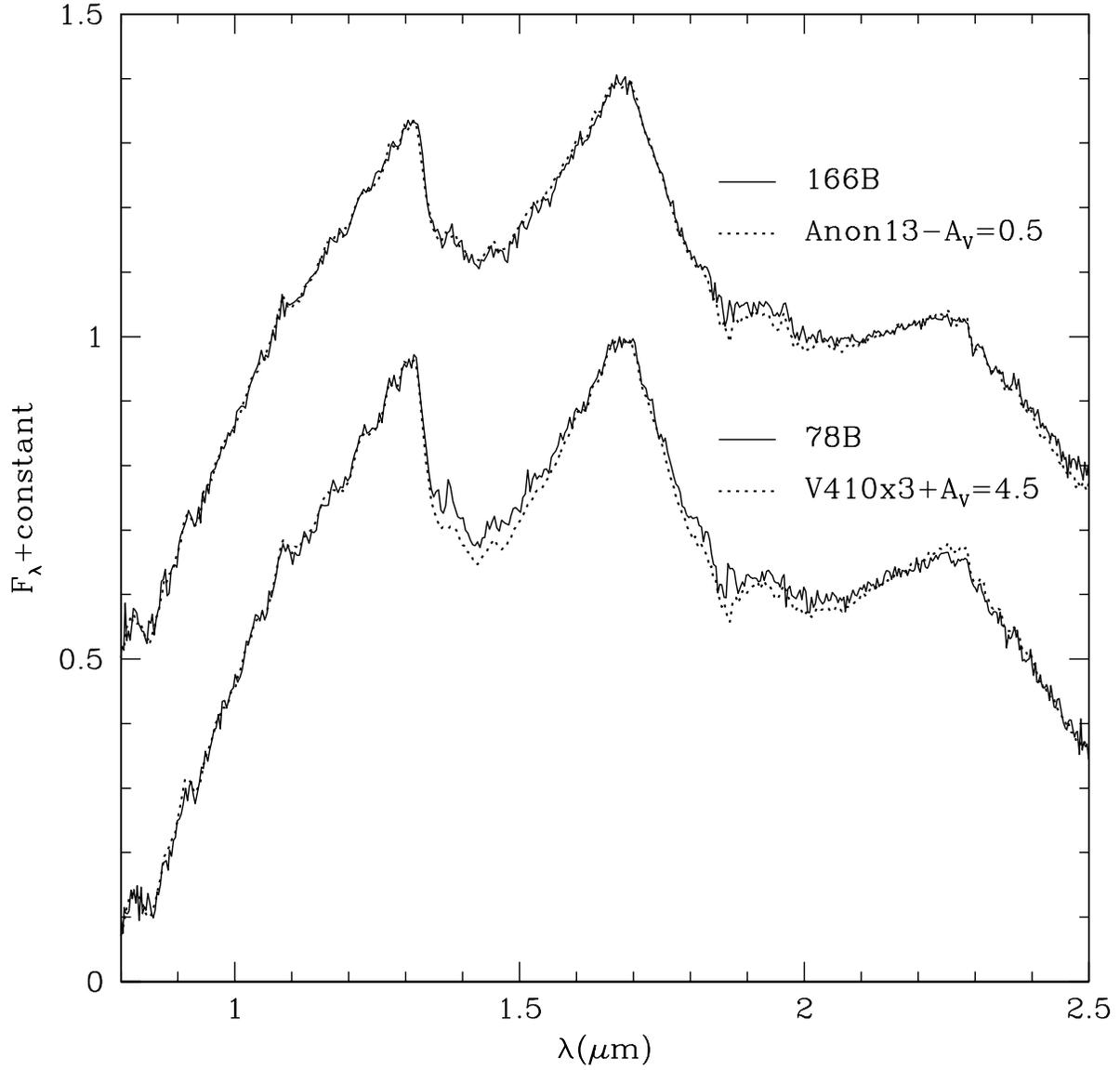}
\caption{
Near-IR spectra of candidate companions 78B and 166B compared to spectra of
V410~X-ray~3 and V410~Anon~13, which are members of the Taurus star-forming
region with optical spectral types of M6 and M5.75, respectively.
The latter have been reddened to match the overall slopes of the spectra of
78B and 166B.
The spectra have a resolution of $R=100$ and are normalized at 1.68~\micron.
}
\label{fig:spex2}
\end{figure}

In \S~\ref{sec:psfpc} and \ref{sec:psfwfc}, 
we measured the photometric detection limits for point sources near stars
in two extreme cases, unsaturated in the PC images and heavily saturated in
the WFC images. We have converted the magnitude differences in these limits 
to mass ratios by applying the theoretical mass-luminosity 
relation for an age of 3~Myr from \citet{bar98} and \citet{cha00} and 
bolometric corrections described in \citet{luh99}. In this conversion, we 
assumed that magnitude differences in F791W are equal to those in the $I$ band. 
We find that the detection limits for unsaturated PC stars in
Figure~\ref{fig:sensitivity} correspond 
to mass ratios of $q\equiv M_2/M_1=0.2$-0.3, 0.1-0.15, and 0.04-0.06 for
$M_1=0.3$-0.1~$M_\odot$ at separations of 0.1, 0.2, and $0\farcs4$, 
respectively. The mass ratio limits
for substellar primaries are similar to those of primaries at 0.1~$M_\odot$, 
except at large separations where limits for the former are higher due to
the smaller magnitude differences between the primary and the sky limit.
For the median age and reddening of members of IC~348, the limiting magnitude
of $m_{791}\sim26$ produced by the sky corresponds to a mass 
of $\sim0.006$~$M_\odot$ according to the evolutionary models of \citet{cha00}.
As noted in \S~\ref{sec:psfwfc}, the detection limits for companions to
unsaturated and lightly saturated stars in the WFC frames are similar to the 
above PC limits, except at double the spatial scale at small separations. 
Meanwhile, for heavily saturated stars with $M_1=0.5$-1.5~$M_\odot$ in the WFC
images, the detection limit at a separation of $0\farcs4$ is $q=0.04$ for
most position angles (Figure~\ref{fig:sensitivity}) and as large as $q=0.4$ 
for angles with strong bleeding. 

\begin{figure}
\plotone{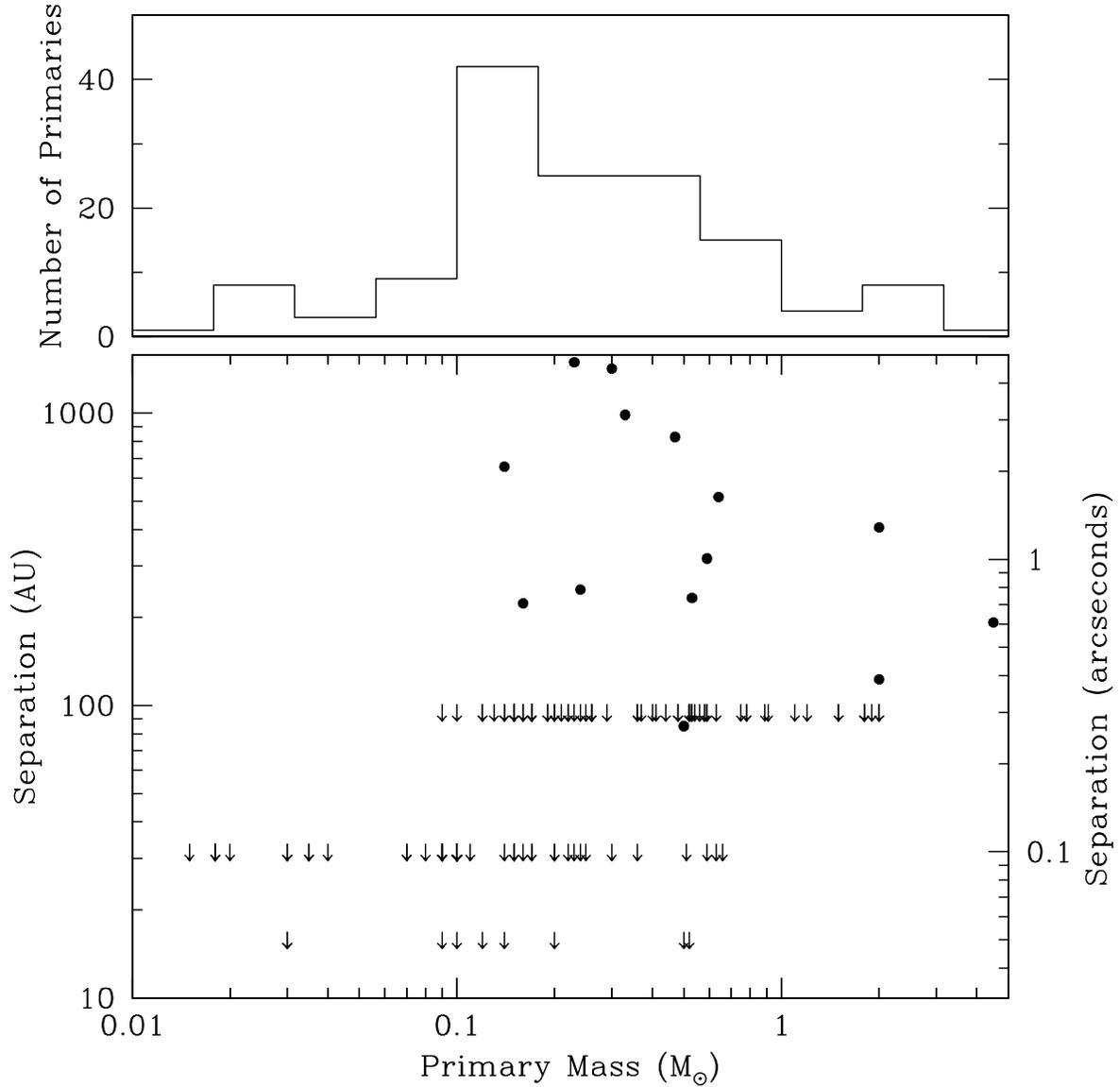}
\caption{
Binary measurements for all known members of the IC~348 cluster within
the WFPC2 frames. {\it Bottom:}
For the 14 members from Table~\ref{tab:pairs} that have known or candidate 
members within $5\arcsec$, we
plot the projected separations of these candidate companions ({\it points}).
We expect that $\sim0$-1 and 2-5 of the 9 and 5 candidate companions at
separations of $\leq2$ and 2-$5\arcsec$ are unrelated cluster members rather
than true companions.
The remaining unresolved members are represented by the upper limits for 
separations of equal-brightness companions ({\it arrows}), which are 
$\sim0\farcs05$ and $0\farcs1$ for unsaturated objects in the PC and WFC, 
respectively, and $\sim0\farcs3$ for saturated stars. Some of the arrows
represent two or more stars with identical positions in this diagram.
{\it Top:} The total number of primaries as a function of mass is shown
with the histogram.
}
\label{fig:pairs}
\end{figure}

To illustrate the results of our companion search, we plot the candidate
companions from Table~\ref{tab:pairs} and the upper separation limits 
for equal-magnitude pairs among the unresolved cluster members in a diagram of
primary mass versus separation in Figure~\ref{fig:pairs}. We also include in 
Figure~\ref{fig:pairs} the total number of primaries within our survey 
as a function of mass. 

\subsection{Solar-Mass Primaries}

Within a range of projected separations of 0.4-$5\arcsec$, 
there are three candidate companions among the 29 solar-mass primaries 
in our sample ($M_1=0.5$-1.5~$M_\odot$).
The least massive candidate is 78B, which has a stellar mass of
$\sim0.1$~$M_\odot$
according to the models of \citet{bar98} and \citet{cha00} (\S~\ref{sec:spex}).
Because these pairs have separations less than $2\arcsec$, they all are 
likely to be true binaries rather than chance alignments of unrelated cluster 
members (\S~\ref{sec:cand}). The resulting binary fraction of stellar companions
is $10^{+9}_{-3}$\%\footnote{The uncertainty is computed in the
manner described by \citet{bur03} and \citet{mz04}}, 
which is consistent with the value of 13\% 
measured for the equivalent range of separations for G dwarfs by \citet{dm91}.
Meanwhile, we detected no objects that are likely to be substellar companions
for this sample of solar-mass stars.
Although the mass detection limit for wide companions in our images 
is much lower than in surveys in the field \citep{giz01a},
we cannot provide useful constraints on the frequency of wide substellar
companions near solar-mass primaries because of the relatively small 
size of our sample for this mass range. 

\subsection{Low-Mass Stellar Primaries}
\label{sec:lowmass}

The number statistics of our
multiplicity measurements are better for the low-mass stars in IC~348
($M_1=0.08$-0.5~$M_\odot$), 85 of which are in our survey.
For this mass range, seven candidate companions appear within projected 
separations of 0.4-$5\arcsec$. 
Six of these candidates are probably above the hydrogen burning limit, 
$\sim2$-4 of which are expected to be true companions rather than unrelated
cluster members (\S~\ref{sec:cand}). After including the Poisson uncertainties, 
the frequency of stellar companions is 2-8\%.
The remaining candidate companion is source 761. This object
is a confirmed late-type cluster member, but the probability of its 
companionship is lower given its relatively wide separation of $4\farcs5$.
If we assume that 761 is a brown dwarf companion, then we derive an upper 
limit of 4\% for the frequency of substellar companions within 
0.4-$5\arcsec$ from low-mass stars in IC~348,
which is complete for $M_2>0.02$~$M_\odot$ for all of the primaries in question.
These frequencies of stellar and substellar companions in IC~348 are roughly
consistent with measurements of $\sim10$\% \citep{fm92} and $\sim1$\% 
\citep{mz04}, respectively, for the same range of projection separations 
near low-mass stars in the field.
Our detection limits reach $M_2\sim0.006$~$M_\odot$ for wider 
separations and less massive primaries. Therefore, in IC~348, we find the same 
absence of widely separated massive planetary companions that has been 
recently reported for young field stars by \citet{mz04}.

\subsection{Brown Dwarf Primaries}

In the census of IC~348 from \citet{luh03}, 23 of the known cluster
members have spectral types later than M6 and thus are likely to be brown 
dwarfs. Our WFPC2 images encompass 15 of these sources, one of which, 761, is 
a candidate companion to a more massive cluster member. For the other 14 brown 
dwarfs, no candidate companions are detected in the WFPC2 data.
If the frequency of wide companions for brown dwarfs in IC~348 is similar to 
that measured in the previous section for the low-mass stars, then we would 
expect to detect at most one companion in this sample of 14 substellar 
primaries. Therefore, the absence of wide binary brown dwarfs in IC~348 does 
not represent a significant difference in the multiplicity of brown dwarfs from
that of the stellar cluster members.
In comparison, several previous surveys for companions to brown dwarfs in 
the field \citep{koe99,mar99,giz03,rei01,clo03,bou03,bur03}, in open clusters
\citep{mar98,mar00,mar03}, and in star-forming regions \citep{neu02}
have also failed to uncover binaries wider than 20~AU, which has been
interpreted as evidence that such systems do not form or are disrupted at very
early stages \citep{bur03}. 
However, the recent discovery a binary brown dwarf with a projected
separation of 240~AU in the Chamaeleon~I star-forming region 
demonstrated that wide binary brown dwarfs indeed do exist \citep{luh04a}.
Multiplicity measurements that reach lower values of $q$ for brown dwarfs in
the field and that consider larger samples for brown dwarfs in star-forming
regions are necessary to determine if the frequency of wide companions varies 
between stars and brown dwarfs and between the field and young clusters.

\subsection{The Brown Dwarf Desert}

According to the binary frequencies from \S~\ref{sec:lowmass}, 
low-mass stars in IC~348 have fewer brown dwarf companions than stellar 
companions at separations of $\sim100$-1000~AU. 
Similar deficiencies in substellar companions have been observed
across a wide range of separations for stars in the field, as illustrated 
in Figure~8 from \citet{mz04}, which compared published frequencies of 
stellar and substellar companions as a function of separation.
The ratio of the frequencies of stellar and substellar companions,
which is between $\sim3$ and 10 at most separations \citep{mz04}, is
comparable to the ratio of the numbers of stars and brown dwarfs
in isolation, which is between $\sim5$ and 8 in star-forming regions
\citep{bri02,mue02,luh03,sle04,luh04b}. 
In other words, the brown dwarf desert -- when defined relative to stars -- 
is present among both companions and isolated objects, which is expected
if they arise from a common formation mechanism (e.g., core fragmentation.)

\section{Conclusions}

We have performed a search for substellar companions to 
young stars and brown dwarfs in the cluster IC~348 with WFPC2 on $HST$.
In the most favorable circumstances in which primaries are unsaturated and 
appear in PC images, the detection limits for companions are
$\Delta m_{791}=0$, 2.5, 4.5, and 6 at separations of 0.05, 0.1, 0.2, and
$0\farcs4$, respectively, which correspond to 
$q=1$, 0.2-0.3, 0.1-0.15, and 0.04-0.06 for
$M_1=0.3$ to $\lesssim0.1$~$M_\odot$ at
15, 30, 60, and 120~AU.
For the other extreme, the limits are $\Delta m_{791}=0$ and 6 at
0.2 and $0\farcs4$, or $q=1$ and 0.04, 
for heavily saturated solar-mass primaries in the WFC images.
At large separations where a primary's PSF no longer dominates the noise, 
the limiting magnitude of $m_{791}\sim26$ corresponds to a mass of 
$\sim0.006$~$M_\odot$ for the median age and reddening of members of IC~348
according to the evolutionary models of \citet{cha00}.

After measuring photometry and astrometry for all point sources appearing in 
the WFPC2 frames, we have selected all objects within 
$5\arcsec$ of known and candidate members of IC~348 and have 
classified them as known members, candidate members, or field stars based
on data from previous studies and their positions in the color-magnitude 
diagram constructed from the WFPC2 photometry. Through this analysis, 
we have identified 14 pairs that consist of two known members or a 
known member and a candidate member, and thus are potential binary systems.
Based on a Monte Carlo simulation of projected separations among the 
members of IC~348, we expect that $\sim2$-6 of these pairs are composed of 
unrelated members of the cluster.
We have presented 0.8-2.5~\micron\ spectra of the two faintest candidate 
companions in our survey, which have projected separations of $0\farcs79$
and $1\farcs64$. A comparison to spectra of optically-classified dwarfs, 
giants, and pre-main-sequence objects indicates that these sources
are young and thus members of IC~348. We have measured spectral types of 
M5.75 and M6 from these data, which imply masses of $\sim0.1$~$M_\odot$ 
according to the models of \citet{cha00} and \citet{bar98}.
No objects that are likely to be substellar companions are detected in this
survey.

We find that the frequencies of stellar and substellar companions 
at 0.4-$5\arcsec$ (120-1600~AU) from low-mass stars ($M_1=0.08$-0.5) in IC~348
are consistent with measurements for field stars \citep{fm92,mz04}, 
in agreement with the previous binary survey of this cluster \citep{duc99}. 
Like companions to field stars across a wide range of separations 
\citep{mb00,mz04}, wide companions to low-mass stars in IC~348 exhibit a low 
abundance of brown dwarfs relative to stars. 
This deficiency in brown dwarfs among companions in IC~348 and the field 
is similar in size ($\sim3$-10) to the deficiency of free-floating brown 
dwarfs ($\sim5$-8) in star-forming regions \citep{luh04b}, which is expected
if companion and isolated brown dwarfs share a common formation mechanism.
Finally, no companions are detected near the 14 primaries within our WFPC2 
images that are likely to be substellar ($>$M6).
Even with this absence of wide binary brown dwarfs, the frequency of wide 
companions for brown dwarfs is statistically consistent with that of the
low-mass stars in IC~348.

\acknowledgments

We thank undergraduate research assistant Carolin Cardamone (Wellesley 2002) 
for help with the data analysis, Bill Joye for improvements to ds9, 
Brian McLeod for providing the image fitting software, and the anonymous
referee for many constructive comments on the manuscript.
We also thank John Rayner for assistance
with the IRTF SpeX observations. This work was 
supported by grant GO-8573 from the Space Telescope Science Institute.
The FLAMINGOS image from \citet{mue03} was obtained under the NOAO
Survey Program ``Towards a Complete Near-Infrared Imaging and Spectroscopic
Survey of Giant Molecular Clouds" (PI: E. Lada) and supported by NSF grants
AST97-3367 and AST02-02976 to the University of Florida. FLAMINGOS was
designed and constructed by the IR instrumentation group
(PI: R. Elston) at the Department of Astronomy at the University of Florida
with support from NSF grant AST97-31180 and Kitt Peak National Observatory.

\begin{deluxetable}{lcccc}
\tablecaption{Summary of Observations \label{tab:log}}
\tablewidth{0pt}
\tablehead{
\colhead{Field} & 
\colhead{$\alpha$(J2000)} & 
\colhead{$\delta$(J2000)} & 
\colhead{Orientation\tablenotemark{a}} &
\colhead{Date}
}
\startdata
POS1   &   03 44 03.59  &  32 02 32.66 &  114.91  &  2001 Jan 15 \\
POS1r  &   03 44 03.59  &  32 02 32.66 &  144.91  &  2001 Jan 19 \\
POS2   &   03 44 19.54  &  32 02 23.89 &  -60.09  &  2000 Sep 05 \\
POS2r  &   03 44 19.54  &  32 02 23.89 &  149.91  &  2001 Dec 14 \\
POS3   &   03 44 32.73  &  32 04 12.27 &  132.91  &  2001 Jan 15 \\
POS3r  &   03 44 32.73  &  32 04 12.27 &  -77.09  &  2001 Sep 03 \\
POS4   &   03 44 21.08  &  32 06 15.35 &  124.91  &  2001 Jan 18 \\
POS4r  &   03 44 21.08  &  32 06 15.35 &  -95.09  &  2001 Oct 24 \\
POS5   &   03 44 45.65  &  32 11 09.82 &  -98.13  &  2000 Oct 22 \\
POS5r  &   03 44 45.65  &  32 11 09.82 &  111.91  &  2001 Jan 05 \\
POS6   &   03 44 39.19  &  32 08 12.34 &  -49.12  &  2001 Jul 16 \\
POS6r  &   03 44 39.19  &  32 08 12.34 &  -85.09  &  2001 Sep 28 \\
POS7   &   03 44 25.70  &  32 09 04.53 &  -61.09  &  2001 Aug 14 \\
POS7r  &   03 44 25.70  &  32 09 04.53 &  -91.09  &  2001 Oct 06 \\
POS8   &   03 44 30.26  &  32 11 34.11 &  -55.09  &  2000 Sep 11 \\
POS8r  &   03 44 30.26  &  32 11 34.11 &  -86.29  &  2001 Sep 30 \\
POS9   &   03 45 01.06  &  32 12 21.75 &  -63.09  &  2000 Sep 11 \\
POS9r  &   03 45 01.06  &  32 12 21.75 &  146.91  &  2000 Dec 25 \\
POS10  &   03 44 45.93  &  32 03 55.59 &  149.91  &  2000 Dec 26 \\
POS10r &   03 44 45.93  &  32 03 55.59 &  -60.09  &  2001 Aug 10 \\
\enddata
\tablenotetext{a}{Position angle of the y axis of the PC.}
\end{deluxetable}

\begin{deluxetable}{lllllllllll}
\tabletypesize{\footnotesize}
\tablewidth{0pt}
\tablecaption{Members of IC 348 in WFPC2 Images \label{tab:mem}}
\tablehead{
\colhead{} &
\colhead{} &
\colhead{} &
\colhead{} &
\colhead{} &
\colhead{Spectral\tablenotemark{b}} &
\colhead{Mass\tablenotemark{c}} \\
\colhead{ID} &
\colhead{$\alpha$(J2000)} &
\colhead{$\delta$(J2000)} &
\colhead{m$_{791}$\tablenotemark{a}} &
\colhead{m$_{791}-$m$_{850}$\tablenotemark{a}} &
\colhead{Type} &
\colhead{($M_{\odot}$)} 
}
\startdata
   1A & 03 44 34.212 &  32 09 46.69 & \nodata & \nodata &     B5 &   4.5   \\
1B & 03 44 34.196 &  32 09 46.12 & \nodata & \nodata & \nodata & \nodata   \\
    5 & 03 44 26.027 &  32 04 30.41 & \nodata & \nodata &     G8 &   2.0   \\
   9A & 03 44 39.178 &  32 09 18.35 & \nodata & \nodata &     G8 &   2.0   \\
9B & 03 44 39.176 &  32 09 18.74 & \nodata & \nodata & \nodata & \nodata   \\
   10 & 03 44 24.664 &  32 10 15.04 & \nodata & \nodata &     F2 &   1.9   \\
 12A & 03 44 31.960 &  32 11 43.84 & \nodata & \nodata &     G0 &   1.4   \\
 12B & 03 44 32.061 &  32 11 43.94 & \nodata & \nodata &     A3 &   2.0   \\
   13 & 03 43 59.641 &  32 01 54.17 &  20.92 &   1.42 &   M0.5 &  0.63   \\
   15 & 03 44 44.716 &  32 04 02.72 & \nodata & \nodata &   M0.5 &  0.63   \\
   16 & 03 44 32.743 &  32 08 37.46 & \nodata & \nodata &     G6 &   1.5   \\
   19 & 03 44 30.820 &  32 09 55.80 & \nodata & \nodata &     A2 &   2.0   \\
   23 & 03 44 38.718 &  32 08 42.05 & \nodata & \nodata &     K3 &   1.8   \\
   26 & 03 43 56.031 &  32 02 13.31 & \nodata & \nodata &     K7 &  0.78   \\
   29 & 03 44 31.533 &  32 08 45.00 & \nodata & \nodata &     K2 &   1.5   \\
   31 & 03 44 18.164 &  32 04 56.98 & \nodata & \nodata &     G1 &   1.8   \\
   32 & 03 44 37.889 &  32 08 04.16 & \nodata & \nodata &     K7 &  0.78   \\
   33 & 03 44 32.586 &  32 08 42.49 & \nodata & \nodata &   M2.5 &  0.52   \\
   35 & 03 44 39.251 &  32 07 35.55 & \nodata & \nodata &     K3 &   1.8   \\
   37 & 03 44 37.991 &  32 03 29.80 & \nodata & \nodata &     K6 &  0.89   \\
   40 & 03 44 29.722 &  32 10 39.84 & \nodata & \nodata &     K8 &  0.75   \\
 42A & 03 44 42.019 &  32 09 00.12 & \nodata & \nodata &  M4.25 &  0.24   \\
 42B & 03 44 42.146 &  32 09 02.20 & \nodata & \nodata &   M2.5 &  0.47   \\
   45 & 03 44 24.290 &  32 10 19.42 & \nodata & \nodata &     K5 &   1.1   \\
   49 & 03 43 57.595 &  32 01 37.57 &  21.31 &   1.33 &   M0.5 &  0.66   \\
   52 & 03 44 43.515 &  32 07 42.98 & \nodata & \nodata &     M1 &  0.59   \\
   58 & 03 44 38.550 &  32 08 00.68 & \nodata & \nodata &  M1.25 &  0.58   \\
   59 & 03 44 40.127 &  32 11 34.34 & \nodata & \nodata &     K2 &   1.2   \\
   62 & 03 44 26.628 &  32 03 58.30 & \nodata & \nodata &  M4.75 &  0.17   \\
   69 & 03 44 27.016 &  32 04 43.62 & \nodata & \nodata &     M1 &  0.59   \\
   71 & 03 44 32.575 &  32 08 55.82 & \nodata & \nodata &     M3 &  0.48   \\
   72 & 03 44 22.575 &  32 01 53.73 & \nodata & \nodata &   M2.5 &  0.52   \\
   74 & 03 44 34.271 &  32 10 49.67 & \nodata & \nodata &     M2 &  0.54   \\
   75 & 03 44 43.784 &  32 10 30.56 & \nodata & \nodata &  M1.25 &  0.58   \\
 78A & 03 44 26.688 &  32 08 20.32 & \nodata & \nodata &   M0.5 &  0.64   \\
   83 & 03 44 37.413 &  32 09 00.93 & \nodata & \nodata &     M1 &  0.59   \\
   88 & 03 44 32.769 &  32 09 15.77 & \nodata & \nodata &  M3.25 &  0.41   \\
   90 & 03 44 33.309 &  32 09 39.62 & \nodata & \nodata &     M2 &  0.53   \\
   91 & 03 44 39.210 &  32 09 44.73 & \nodata & \nodata &     M2 &  0.53   \\
  92A & 03 44 23.672 &  32 06 46.56 & \nodata & \nodata &   M2.5 &  0.50   \\
92B & 03 44 23.668 &  32 06 46.83 & \nodata & \nodata & \nodata & \nodata   \\
   97 & 03 44 25.559 &  32 06 16.95 & \nodata & \nodata &  M2.25 &  0.53   \\
   98 & 03 44 38.615 &  32 05 06.45 & \nodata & \nodata &     M4 &  0.29   \\
  99A & 03 44 19.254 &  32 07 34.68 & \nodata & \nodata &  M3.75 &  0.33   \\
  99B & 03 44 19.021 &  32 07 35.69 & \nodata & \nodata &  M5.25 &  0.13   \\
  103 & 03 44 44.594 &  32 08 12.66 & \nodata & \nodata &     M2 &  0.53   \\
  108 & 03 44 38.700 &  32 08 56.74 & \nodata & \nodata &  M3.25 &  0.37   \\
  110 & 03 44 37.398 &  32 12 24.26 & \nodata & \nodata &     M2 &  0.52   \\
  113 & 03 44 37.193 &  32 09 16.10 & \nodata & \nodata &     K6 &  0.91   \\
  115 & 03 44 30.003 &  32 09 21.07 &  18.65 &   1.06 &   M2.5 &  0.51   \\
  116 & 03 44 21.559 &  32 10 17.38 & \nodata & \nodata &   M1.5 &  0.56   \\
  119 & 03 44 21.256 &  32 05 02.44 & \nodata & \nodata &   M2.5 &  0.48   \\
  123 & 03 44 24.567 &  32 03 57.12 &  17.10 &  \nodata &     M1 &  0.59   \\
  125 & 03 44 21.664 &  32 06 24.80 & \nodata & \nodata &  M2.75 &  0.44   \\
  133 & 03 44 41.743 &  32 12 02.41 & \nodata & \nodata &     M5 &  0.15   \\
  139 & 03 44 25.308 &  32 10 12.65 & \nodata & \nodata &  M4.75 &  0.17   \\
  145 & 03 44 41.308 &  32 10 25.30 & \nodata & \nodata &  M4.75 &  0.17   \\
  151 & 03 44 34.830 &  32 11 18.00 & \nodata & \nodata &     M2 &  0.52   \\
  153 & 03 44 42.773 &  32 08 33.86 & \nodata & \nodata &  M4.75 &  0.17   \\
  154 & 03 44 37.788 &  32 12 18.21 & \nodata & \nodata &   M4.5 &  0.23   \\
  158 & 03 44 40.164 &  32 09 13.00 & \nodata & \nodata &     M5 &  0.15   \\
  159 & 03 44 47.624 &  32 10 55.79 & \nodata & \nodata &  M4.25 &  0.26   \\
  160 & 03 44 02.591 &  32 01 35.07 & \nodata & \nodata &  M4.75 &  0.20   \\
  165 & 03 44 35.465 &  32 08 56.53 & \nodata & \nodata &  M5.25 &  0.14   \\
  166A & 03 44 42.581 &  32 10 02.50 &  18.60 &   1.20 &  M4.25 &  0.24   \\
  167 & 03 44 41.179 &  32 10 10.18 & \nodata & \nodata &     M3 &  0.36   \\
  168 & 03 44 31.351 &  32 10 46.89 & \nodata & \nodata &  M4.25 &  0.24   \\
  171 & 03 44 44.851 &  32 11 05.76 & \nodata & \nodata &  M2.75 &  0.40   \\
  175 & 03 44 49.795 &  32 03 34.21 & \nodata & \nodata &   M4.5 &  0.22   \\
  186 & 03 44 46.319 &  32 11 16.75 &  18.80 &   1.03 &     M2 &  0.52   \\
  192 & 03 44 23.641 &  32 01 52.69 &  20.01 &   1.28 &   M4.5 &  0.23   \\
  193 & 03 44 38.012 &  32 11 37.10 & \nodata & \nodata &     M4 &  0.26   \\
  194 & 03 44 27.252 &  32 10 37.28 & \nodata & \nodata &  M4.75 &  0.20   \\
  201 & 03 45 01.485 &  32 12 29.06 & \nodata & \nodata &     M4 &  0.26   \\
  210 & 03 44 20.020 &  32 06 45.57 & \nodata & \nodata &   M3.5 &  0.30   \\
  217 & 03 44 43.055 &  32 10 15.29 & \nodata & \nodata &     M5 &  0.17   \\
  218 & 03 44 44.663 &  32 07 30.31 & \nodata & \nodata &  M5.25 &  0.15   \\
  221 & 03 44 40.255 &  32 09 33.23 &  18.29 &   0.95 &   M4.5 &  0.20   \\
  226 & 03 44 31.425 &  32 11 29.52 &  18.18 &   1.14 &  M5.25 &  0.16   \\
  237 & 03 44 23.570 &  32 09 33.98 & \nodata & \nodata &     M5 &  0.17   \\
  240 & 03 44 52.101 &  32 04 46.86 & \nodata & \nodata &     M4 &  0.25   \\
  241 & 03 44 59.837 &  32 13 32.20 & \nodata & \nodata &   M4.5 &  0.22   \\
  242 & 03 44 32.810 &  32 04 13.11 & \nodata & \nodata &     M5 &  0.17   \\
  248 & 03 44 35.949 &  32 09 24.29 & \nodata & \nodata &  M5.25 &  0.15   \\
  252 & 03 44 29.118 &  32 07 57.36 & \nodata & \nodata &   M4.5 &  0.21   \\
  255 & 03 44 35.701 &  32 04 52.71 & \nodata & \nodata &  M5.75 &  0.12   \\
 259A & 03 44 03.651 &  32 02 35.12 &  17.91 &   0.72 &     M5 &  0.14   \\
 259B & 03 44 03.620 &  32 02 33.08 &  18.05 &   0.75 &     M5 &  0.14   \\
  266 & 03 44 18.270 &  32 07 32.46 & \nodata & \nodata &  M4.75 &  0.19   \\
  277 & 03 44 39.444 &  32 10 08.17 & \nodata & \nodata &     M5 &  0.16   \\
  278 & 03 44 31.031 &  32 05 45.92 & \nodata & \nodata &   M5.5 &  0.13   \\
  285 & 03 44 31.853 &  32 12 44.17 &  19.67 &   1.07 &   M4.5 &  0.22   \\
  287 & 03 44 41.115 &  32 08 07.49 &  19.33 &   1.05 &  M5.25 &  0.15   \\
  294 & 03 44 24.581 &  32 10 02.93 & \nodata & \nodata &   M4.5 &  0.16   \\
  295 & 03 44 29.511 &  32 04 04.43 &  18.60 &   0.90 &     M5 &  0.17   \\
  300 & 03 44 38.980 &  32 03 19.80 & \nodata & \nodata &     M5 &  0.15   \\
  301 & 03 44 22.693 &  32 01 42.31 &  20.13 &   1.12 &  M4.75 &  0.17   \\
  302 & 03 44 20.276 &  32 05 43.69 & \nodata & \nodata &  M4.75 &  0.19   \\
  308 & 03 44 21.218 &  32 01 14.50 &  22.43 &   1.41 &     M4 &  0.24   \\
  309 & 03 44 31.340 &  32 09 29.20 & \nodata & \nodata &     M3 &  0.36   \\
  314 & 03 44 22.552 &  32 01 27.74 &  20.16 &   1.13 &     M5 &  0.15   \\
  319 & 03 45 01.003 &  32 12 22.48 & \nodata & \nodata &   M5.5 &  0.12   \\
  322 & 03 44 19.580 &  32 02 24.89 &  18.79 &   0.81 &  M4.25 &  0.20   \\
  324 & 03 44 45.225 &  32 10 55.94 &  18.66 &   0.90 &  M5.75 &  0.10   \\
  325 & 03 44 30.054 &  32 08 48.84 &  18.91 &   0.98 &     M6 & 0.090   \\
  334 & 03 44 26.663 &  32 02 36.39 & \nodata & \nodata &  M5.75 &  0.10   \\
  335 & 03 44 44.252 &  32 08 47.41 &  18.68 &   0.88 &  M5.75 &  0.10   \\
  336 & 03 44 32.364 &  32 03 27.32 &  19.01 &   0.96 &   M5.5 &  0.10   \\
  342 & 03 44 41.316 &  32 04 53.48 & \nodata & \nodata &     M5 &  0.15   \\
  347 & 03 44 27.284 &  32 07 17.68 &  18.21 &   0.73 &  M4.75 &  0.15   \\
  350 & 03 44 19.181 &  32 05 59.80 &  18.36 &   0.87 &  M5.75 &  0.10   \\
  351 & 03 44 25.749 &  32 09 06.01 &  18.90 &   0.96 &   M5.5 &  0.12   \\
  353 & 03 44 38.155 &  32 10 21.59 & \nodata & \nodata &     M6 & 0.090   \\
  355 & 03 44 39.199 &  32 08 13.93 &  19.51 &   1.21 &     M8 & 0.030   \\
  360 & 03 44 43.720 &  32 10 48.10 & \nodata & \nodata &  M4.75 &  0.14   \\
  366 & 03 44 35.027 &  32 08 57.54 &  18.77 &   0.94 &  M4.75 &  0.14   \\
  373 & 03 44 27.985 &  32 05 19.64 &  18.53 &   0.75 &   M5.5 &  0.10   \\
  382 & 03 44 30.943 &  32 02 44.16 &  20.36 &   1.25 &   M5.5 &  0.11   \\
  385 & 03 44 28.871 &  32 04 22.87 &  19.28 &   0.91 &  M5.75 & 0.090   \\
  391 & 03 44 46.589 &  32 09 01.86 &  20.01 &   1.07 &  M5.75 & 0.090   \\
  405 & 03 44 21.163 &  32 06 16.56 &  19.69 &   1.13 &     M8 & 0.030   \\
  410 & 03 44 37.557 &  32 11 55.81 &  23.38 &   1.68 &     M4 &  0.25   \\
  413 & 03 44 45.646 &  32 11 10.86 &  18.62 &   0.68 &  M4.75 &  0.14   \\
  414 & 03 44 44.293 &  32 10 36.90 &  19.09 &   0.78 &  M5.25 &  0.10   \\
  415 & 03 44 29.983 &  32 09 39.48 &  19.64 &   1.05 &   M6.5 & 0.070   \\
  432 & 03 44 45.950 &  32 03 56.80 &  19.52 &   1.01 &  M5.75 & 0.090   \\
  435 & 03 44 30.278 &  32 11 35.27 &  20.17 &   0.53 &  M2.25 &  0.50   \\
  454 & 03 44 41.575 &  32 10 39.47 &  19.15 &   0.73 &  M5.75 & 0.080   \\
  462 & 03 44 24.449 &  32 01 43.69 &  20.59 &   1.19 &     M3 &  0.36   \\
  478 & 03 44 35.940 &  32 11 17.51 &  20.04 &   0.97 &  M6.25 & 0.070   \\
  603 & 03 44 33.414 &  32 10 31.55 &  21.47 &   1.25 &   M8.5 & 0.020   \\
  611 & 03 44 30.372 &  32 09 44.57 &  21.01 &   1.10 &     M8 & 0.035   \\
  613 & 03 44 26.885 &  32 09 26.21 &  21.32 &   1.22 &  M8.25 & 0.030   \\
  622 & 03 44 31.333 &  32 08 11.45 &  21.52 &   0.96 &     M6 &  0.10   \\
  624 & 03 44 26.367 &  32 08 09.94 &  23.18 &   1.49 &     M9 & 0.018   \\
  690 & 03 44 36.381 &  32 03 05.42 &  21.39 &   1.36 &  M8.75 & 0.018   \\
  703 & 03 44 36.618 &  32 03 44.23 &  21.60 &   1.35 &     M8 & 0.035   \\
  705 & 03 44 27.169 &  32 03 46.60 &  22.35 &   1.34 &     M9 & 0.015   \\
  725 & 03 44 33.699 &  32 05 20.67 &  22.23 &   0.94 &     M6 &  0.10   \\
  761 & 03 44 19.666 &  32 06 45.93 &  21.47 &   1.18 &     M7 & 0.060   \\
  906 & 03 45 03.606 &  32 12 13.95 &  21.75 &   1.15 &  M8.25 & 0.040   \\
  935 & 03 44 26.917 &  32 12 50.66 &  20.88 &   1.16 &  M8.25 & 0.030   \\
 1434 & 03 44 22.983 &  32 07 18.99 &  22.38 &   0.82 &     M6 &  0.10   \\
 1684 & 03 44 23.294 &  32 01 54.43 &  18.80 &   0.88 &  M5.75 & 0.090   \\
 1868 & 03 45 01.588 &  32 13 17.03 &  18.43 &   0.93 &     M4 &  0.23   \\
 4044 & 03 44 16.176 &  32 05 40.96 &  22.99 &   1.35 &     M9 & 0.018   \\
  362 & 03 44 42.304 &  32 12 28.30 &  22.72 &   1.74 &     M5 &  0.16   \\
 1477 & 03 44 36.249 &  32 13 04.55 &  23.07 &   1.53 &     M6 &  0.09   \\
  202 & 03 44 34.276 &  32 12 40.71 &  21.72 &   1.48 &   M3.5 &   0.3   \\
  297 & 03 44 33.212 &  32 12 57.45 &  20.91 &   1.40 &   M4.5 &   0.2   \\
\enddata
\tablenotetext{a}{Stars without measurements are saturated.}
\tablenotetext{b}{Spectral types adopted by \citet{luh03} and \citet{luh05}.}
\tablenotetext{c}{Masses derived by \citet{luh03}.}
\end{deluxetable}

\begin{deluxetable}{lllllllllllll}
\tabletypesize{\scriptsize}
\rotate
\tablewidth{0pt}
\tablecaption{Pairs in WFPC2 Images of IC 348 \label{tab:pairs}}
\tablehead{
\colhead{ID(A)} &
\colhead{$\alpha$(J2000)} &
\colhead{$\delta$(J2000)} &
\colhead{m$_{791}$} &
\colhead{m$_{791}-$m$_{850}$} &
\colhead{Status\tablenotemark{a}} &
\colhead{ID(B)} &
\colhead{$\alpha$(J2000)} &
\colhead{$\delta$(J2000)} &
\colhead{m$_{791}$} &
\colhead{m$_{791}-$m$_{850}$} &
\colhead{Status\tablenotemark{a}} &
\colhead{Sep\tablenotemark{b}} 
}
\startdata
   1A & 03 44 34.212 &  32 09 46.69 & \nodata & \nodata &      m & 1B & 03 44 34.196 &  32 09 46.12 & \nodata & \nodata &      m &   0.61   \\
   9A & 03 44 39.178 &  32 09 18.35 & \nodata & \nodata &      m & 9B & 03 44 39.176 &  32 09 18.74 & \nodata & \nodata &      m &   0.39   \\
   9A & 03 44 39.178 &  32 09 18.35 & \nodata & \nodata &      m & 30085 & 03 44 39.238 &  32 09 15.44 &  24.70 &   1.10 &      f &   3.01   \\
 12A & 03 44 31.960 &  32 11 43.84 & \nodata & \nodata &      m &  12B & 03 44 32.061 &  32 11 43.94 & \nodata & \nodata &      m &   1.29   \\
   15 & 03 44 44.716 &  32 04 02.72 & \nodata & \nodata &      m & 30169 & 03 44 44.396 &  32 04 04.10 &  23.27 &   0.58 &      f &   4.30   \\
   15 & 03 44 44.716 &  32 04 02.72 & \nodata & \nodata &      m & 30170 & 03 44 44.385 &  32 04 04.07 &  23.18 &   0.62 &      f &   4.42   \\
   40 & 03 44 29.722 &  32 10 39.84 & \nodata & \nodata &      m &  6009 & 03 44 29.922 &  32 10 39.23 &  22.46 &   0.74 &      f &   2.61   \\
 42A & 03 44 42.019 &  32 09 00.12 & \nodata & \nodata &      m &  42B & 03 44 42.146 &  32 09 02.20 & \nodata & \nodata &      m &   2.63   \\
   52 & 03 44 43.515 &  32 07 42.98 & \nodata & \nodata &      m & 30095 & 03 44 43.548 &  32 07 42.06 &  18.38 &   0.95 & c(0.2) &   1.01   \\
   75 & 03 44 43.784 &  32 10 30.56 & \nodata & \nodata &      m & 30187 & 03 44 43.840 &  32 10 31.09 &  22.70 &   0.80 &      f &   0.89   \\
 78A & 03 44 26.688 &  32 08 20.32 & \nodata & \nodata &      m &  78B & 03 44 26.561 &  32 08 20.64 &  20.77 &   1.35 & c(0.1)\tablenotemark{c} &   1.64   \\
  92A & 03 44 23.672 &  32 06 46.56 & \nodata & \nodata &      m & 92B & 03 44 23.668 &  32 06 46.83 & \nodata & \nodata &      m &   0.27   \\
   97 & 03 44 25.559 &  32 06 16.95 & \nodata & \nodata &      m & 30055 & 03 44 25.503 &  32 06 17.14 &  19.23 &   1.03 & c(0.2) &   0.74   \\
  99A & 03 44 19.254 &  32 07 34.68 & \nodata & \nodata &      m &  99B  & 03 44 19.021 &  32 07 35.69 & \nodata & \nodata &      m &   3.13   \\
  123 & 03 44 24.567 &  32 03 57.12 &  17.10 & \nodata &      m & 30188 & 03 44 24.498 &  32 03 58.73 &  $\sim26.8$ & \nodata &      ? &   1.83   \\
  165 & 03 44 35.465 &  32 08 56.53 & \nodata & \nodata &      m & 30099 & 03 44 35.546 &  32 08 55.64 &  21.59 &   1.02 &     f? &   1.36   \\
  166A & 03 44 42.581 &  32 10 02.50 &  18.60 &   1.20 &      m & 166B & 03 44 42.591 &  32 10 03.28 &  20.38 &   1.35 & c(0.1)\tablenotemark{c} &   0.79   \\
  166A & 03 44 42.581 &  32 10 02.50 &  18.60 &   1.20 &      m &  3005 & 03 44 42.274 &  32 10 01.58 &  23.74 &   0.53 &      f &   4.01   \\
  210 & 03 44 20.020 &  32 06 45.57 & \nodata & \nodata &      m &   761 & 03 44 19.666 &  32 06 45.93 &  21.47 &   1.18 &      m &   4.51   \\
  226 & 03 44 31.425 &  32 11 29.52 &  18.18 &   1.14 &      m & 30114 & 03 44 31.395 &  32 11 28.92 &  18.57 &   0.87 & c(0.1) &   0.71   \\
  242 & 03 44 32.810 &  32 04 13.11 & \nodata & \nodata &      m & 30018 & 03 44 32.712 &  32 04 14.96 &  22.56 &   0.68 &      f &   2.23   \\
  259A & 03 44 03.651 &  32 02 35.12 &  17.91 &   0.72 &      m & 259B & 03 44 03.620 &  32 02 33.08 &  18.05 &   0.75 &      m &   2.08   \\
  266 & 03 44 18.270 &  32 07 32.46 & \nodata & \nodata &      m & 30048 & 03 44 18.115 &  32 07 27.98 &  25.59 &   1.08 &      f &   4.89   \\
  295 & 03 44 29.511 &  32 04 04.43 &  18.60 &   0.90 &      m & 22316 & 03 44 29.872 &  32 04 03.03 &  24.35 &   0.97 &      f &   4.80   \\
  385 & 03 44 28.871 &  32 04 22.87 &  19.28 &   0.91 &      m &  4028 & 03 44 28.597 &  32 04 24.33 &  23.86 &   0.87 &      f &   3.78   \\
  432 & 03 44 45.950 &  32 03 56.80 &  19.52 &   1.01 &      m &   707 & 03 44 45.978 &  32 03 53.49 &  21.49 &   0.30 &      f &   3.33   \\
  596 & 03 44 35.164 &  32 11 05.25 &  25.64 &   1.57 &      c & 30115 & 03 44 35.400 &  32 11 04.23 &  25.19 &   1.01 &      f &   3.16   \\
  613 & 03 44 26.885 &  32 09 26.21 &  21.32 &   1.22 &      m & 30101 & 03 44 26.735 &  32 09 30.62 &  24.99 &   0.87 &      f &   4.80   \\
 1434 & 03 44 22.983 &  32 07 18.99 &  22.38 &   0.82 &      m & 30052 & 03 44 23.176 &  32 07 21.98 &  25.56 &   0.84 &      f &   3.87   \\
 1684 & 03 44 23.294 &  32 01 54.43 &  18.80 &   0.88 &      m &   192 & 03 44 23.641 &  32 01 52.69 &  20.01 &   1.28 &      m &   4.74   \\
 1868 & 03 45 01.588 &  32 13 17.03 &  18.43 &   0.93 &      m &  3069 & 03 45 01.477 &  32 13 15.56 &  21.96 &   0.52 &      f &   2.04   \\
\enddata
\tablenotetext{a}{m=member of IC~348 \citep{luh03}; 
c or f=candidate member or field star by the color-magnitude diagram 
in Figure~\ref{fig:iz}. For the candidate members, mass estimates in
solar masses are included in parenthesis.}
\tablenotetext{b}{Separation between stars A and B in arcseconds.}
\tablenotetext{c}{Confirmed as a cluster member with spectroscopy in this work.}
\end{deluxetable}

\clearpage

\clearpage

\clearpage

\clearpage

\clearpage

\clearpage

\clearpage

\clearpage

\clearpage

\clearpage

\end{document}